\documentclass[journal,twocolumn]{IEEEtran}

\usepackage[ruled,linesnumbered]{algorithm2e}
\usepackage{enumerate}
\usepackage{epsfig}
\usepackage{graphicx}
\usepackage{amsmath}
\usepackage{amssymb}
\usepackage{url}
\usepackage{subfigure}
\usepackage{wrapfig}
\usepackage[dvips]{color}
\usepackage{paralist}
\usepackage{booktabs} 
\usepackage{multirow}
\usepackage{tabularx}
\usepackage{paralist}

\usepackage{blindtext,graphicx}
\usepackage[absolute]{textpos}

\usepackage{url}

\newcommand{\newmaterial}[1]{\textcolor{black}{{#1}}}

\begin{document}
\begin{textblock}{12}(2,0.2)
\noindent\small The final version of this article appears in IEEE Transactions on Mobile Computing,  and is accessible via http://ieeexplore.ieee.org/abstract/document/7887755/
\end{textblock}

\title{Video Liveness for Citizen Journalism: Attacks and Defenses}

\author{Mahmudur~Rahman, Mozhgan~Azimpourkivi, Umut~Topkara, Bogdan~Carbunar
   \IEEEcompsocitemizethanks{
      \IEEEcompsocthanksitem Mahmudur Rahman is with IBM Watson, Raleigh, NC, USA. E-mail: mrahman.fiu@gmail.com
      \IEEEcompsocthanksitem Mozhgan Azimpourkivi and Bogdan
Carbunar are with the School of Computing and Information Sciences at the
Florida International University, Miami, FL, USA.  E-mail: \{mozhganaz,carbunar\}@cs.fiu.edu
      \IEEEcompsocthanksitem Umut Topkara is with JW Player
in New York, NY, USA. E-mail: topkara@gmail.com
      \IEEEcompsocthanksitem A preliminary version of this article has appeared in ACM WiSec 2015.
   }
}

\maketitle

\begin{abstract}
The impact of citizen journalism raises important video integrity and
credibility issues. In this article, we introduce Vamos, the first user
transparent video ``liveness'' verification solution based on video motion,
that accommodates the full range of camera movements, and supports videos of
arbitrary length.  Vamos uses the agreement between video motion and camera
movement to corroborate the video authenticity. Vamos can be integrated into any
mobile video capture application without requiring special user training.  We
develop novel attacks that target liveness verification solutions. The attacks
leverage both fully automated algorithms and trained human experts. We
introduce the concept of video motion categories to annotate the camera and
user motion characteristics of arbitrary videos.  We show that the performance
of Vamos depends on the video motion category.  Even though Vamos uses
motion as a basis for verification, we observe a surprising and seemingly
counter-intuitive resilience against attacks performed on relatively
``stationary'' video chunks, which turn out to contain hard-to-imitate
involuntary movements. We show that overall the accuracy of Vamos on the task
of verifying whole length videos exceeds 93\% against the new attacks.
\end{abstract}

\section{Introduction}

The citizen journalism revolution, enabled by advances in mobile and social
technologies, transforms information consumers into collectors and
disseminators of news.  Major news outlets have started to fill out
professional journalistic gaps with videos shot on mobile devices.  Examples
range from videos of conflicts in areas with limited professional journalism
representation (e.g., Syria, Ukraine) to spontaneous events (e.g., tsunamis,
earthquakes, meteorite landings, authority abuse). Such videos are often
distributed through sites such as CNN's iReport~\cite{iReport}, NBC's
Stringwire~\cite{Stringwire} and CitizenTube~\cite{CitizenTube}.

\begin{figure}
\centering
\includegraphics[width=0.43\textwidth]{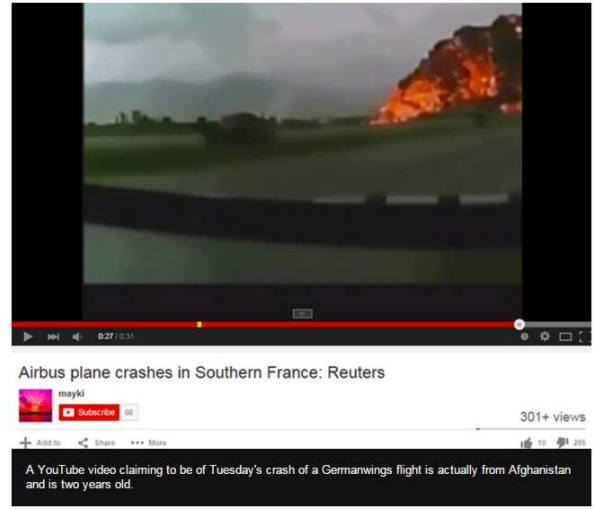}
\caption{On BBC: Video shot in Afghanistan claimed to be of the
Germanwings crash.
\label{fig:fake.video.bbc}}
\vspace{-10pt}
\end{figure}

The increasing popularity of citizen journalism is starting however to raise
important questions concerning the credibility of impactful videos (see
e.g.,~\cite{CitizenEvidenceLab,WitnessOrg,DisturbingFake,Guardian}).
The potential impact of such videos, coupled with the use of
financial incentives, can motivate workers to fabricate data. The media abounds
with examples of fraudulent videos and
images~\cite{Fake.Germanwings,Fake.MH370,Rohingya,Fake.Syrian.Hero}, often
captured at different locations than claimed (see e.g.,
Figure~\ref{fig:fake.video.bbc}).

Videos from other sources can be copied, projected and recaptured, cut and
stitched before being uploaded as genuine on social media sites.  For instance,
plagiarized videos with fabricated location and time stamps can be created
through ``projection'' attacks: the attacker uses specialized
apps~\cite{Android.FakeGPS,Android.FakeGPS1,Android.FakeGPS2} to set the GPS
position of the device to a desired location, then uses the device to shoot a
projected version of the target video.

Citizen Evidence Lab~\cite{CitizenEvidenceLab} and Witness.Org~\cite{Witness}
provide tutorials to train the public to create and to assess citizen reports,
including those captured with mobile devices.
%
%
InformaCam~\cite{InformaCam} provides mechanisms to ensure that the media was
captured by a specific device at a certain location and time.  InformaCam is
however ineffective against adversaries that capture projected videos: while
the resulting videos are fraudulent, they \textit{have} been shot with the
claimed device, at the claimed location and time. While manual verifications
can detect projection attacks, they do not scale well to the high number of
videos on social media sites.

To address this problem, we exploit the observation that for plagiarized
videos, the motion encoded in the video stream is likely inconsistent with the
motion from the inertial sensor streams (e.g., accelerometer) of the device.
Movee~\cite{RTC13}, a video liveness verification solution that uses this
principle, has important weaknesses: i) it is not user transparent to 
the extent that it imposes an explicit verification step on users, ii) it severely 
limits the movements in the verification step to one of four pan movements, and 
iii) it is vulnerable to ``stitch'' attacks in which the attacker creates a 
fraudulent video by first live recording a genuine video and then pointing the camera
to a pre-recorded target video.

In this paper, we introduce Vamos, a Video Accreditation through Motion
Signatures system. Vamos provides liveness verifications for videos of
arbitrary length.
%
%
Vamos is completely transparent to the users; it requires no special user
interaction, nor change in user behavior.

Instead of enforcing an initial verification step, Vamos uses the entire video
and acceleration stream for verification purposes: It divides the video and
acceleration data into fixed length chunks. It then classifies each chunk and
uses the results, along with a suite of novel features that we introduce, to
classify the entire sample. Vamos does not impose a dominant motion direction,
thus, does not constrain the user movements. Instead, Vamos verifies the
liveness of the video by extracting features from \textit{all} the directions
of movement, from both the video and acceleration streams.

Removed video length and movement constraints provide additional flexibility
for attackers to create fraudulent videos. In order to study the security of
the new unconstrained setting, we i) propose a novel, motion based video
classification system, ii) introduce a suite of human centric and fully
automated attacks that target sensor based video liveness verification systems,
and iii) show experimental evidence on a wide range of data collected through
user studies and from public sources.

The attacks we introduce seek to produce accelerometer readings that enable the
attackers to thwart Vamos and claim the production of plagiarized videos, on
their mobile devices. While some of the attacks leverage accelerometer data
produced on the attacker device, several attacks enable the attacker to
fabricate accelerometer readings of their choice.

\newmaterial{
Since Vamos leverages accelerometer data simultaneously captured on the same
device with the video, it can only validate videos recorded using the Vamos
``client'' app, see Section~\ref{sec:model}.}

To evaluate Vamos, we have collected 150 citizen journalism videos from YouTube
and performed a user study to collect 160 free-form videos and corresponding
acceleration samples. Our experiments show that Vamos improves on the free-form
video motion verification accuracy of Movee by more than $15\%$ in the domain
of $6$ second {\it cluster} and i{\it sandwich} attack videos, and by more than
$30\%$ in the domain of whole length {\it stitch} attack videos (see
Section~\ref{sec:model} for the attack description). We show that Vamos
performs unexpectedly well on a suite of {\it mirror} attack variants, that
fabricate acceleration readings based on video motion streams.

We posit that the success rate of the attacks depends on the type
of motions encoded in the video. Experiments with 6s chunks extracted from the
free-form videos confirm (through $\chi^2$ and Fisher's exact tests) that the
classification performance of Vamos depends on the video category.
We show that the proposed motion based video classification can be
used to predict the accuracy of Vamos on videos for which we currently lack
associated sensor streams (e.g., YouTube videos).  To summarize, this paper
makes the following contributions:

\begin{compactitem}

\item
{\bf Targeted attacks}. Introduce a sensor based attack model and develop
manual and automatic attacks targeted against video verification mechanisms.
[\S~\ref{sec:model}].

\item
{\bf Video motion classification}. Introduce a novel classification of mobile
videos [\S~\ref{sec:category}].

\item
{\bf Vamos}. Develop a video liveness verification solution to detect
fraudulent video and inertial sensor chunks that encode arbitrary motions.
Introduce Vamos, a system that detects fraudulent video and accelerometer
streams of arbitrary length, and is resilient to powerful attacks
[\S~\ref{sec:vamos}].

\item
{\bf Video data collection}. Collect datasets of free-form and citizen
journalism videos [\S~\ref{sec:data}].

\item
{\bf Extensive evaluation}.
Show that the performance of Vamos is dependent on the video motion
classification [\S~\ref{sec:eval:clvamos}]. Predict the classification of Vamos
on sensor-less citizen journalism videos. Perform a realistic
evaluation of Vamos on a mixture of attacks and video categories, including on
samples belonging to previously unseen attacks and categories.

\end{compactitem}

\noindent
In our experiments we observe a surprising and seemingly counter-intuitive
resilience of Vamos against attacks on ``stationary'' video chunks. We argue
this is due to the ability of Vamos to exploit the involuntary user hand shakes
that occur during video capture sessions. Furthermore, our experiments show
that Vamos differentiates between genuine and fraudulent video and acceleration
samples of unconstrained length and motion, with an accuracy that exceeds 93\%.

\section{The Problem, Motivation\\ and Related Work}
\label{sec:related}


The problem of verifying the authenticity of videos uploaded to a social media
site (e.g., from conflicts in Syria, Ukraine or Venezuela) is paramount to the
ability to use such videos as evidence or trusted sources for journalism.
Citizen Evidence Lab observes that it is common occurrence during complex
emergencies and natural disasters for old pictures and videos to be recycled as
new online, and go viral due to uncritical re-sharing through social
networks~\cite{CitizenEvidenceLab}. 

This problem has several dimensions, that include assessing the location and
time of capture, or the content of the video. For instance, CitizenEvidenceLab
provides tutorials to train the public to asses citizen videos from YouTube.
%
%
It also provides
tools (e.g., YouTube DataViewer) to enable users to extract the exact local
upload time, all thumbnails and audio only from YouTube videos.


Witness.Org~\cite{Witness} is an organization that
trains and supports people using video in their fight for human rights. It
provides a set of rules
%
%
that enable concerned citizens to safely capture quality videos that witness
important events.

InformaCam~\cite{InformaCam} leverages the unique noise of the device camera to
sign content it produces, along with the output of other sensors (e.g., GPS).
This enables InformaCam to authenticate that content has been produced with a
certain camera. InformaCam assumes that the sensor data is valid and has not
been fabricated, and is vulnerable to plagiarism attacks where the attacker
points the camera to a projected video.

In this paper we focus on liveness verifications: verify that the video was
captured on a mobile device, and has not been fabricated using material from
other sources.

Movee~\cite{RTC13} is a liveness verification solution that imposes a 6 second
verification step on video capturing experiences: before being allowed to shoot
the desired video, the user is presented with a target (bullseye) symbol
displayed randomly either on the top, bottom, left or the right side of the
screen. The user needs to align the center of the screen to the bullseye, by
moving the device in its direction.

Movee uses the correspondence between the motion sensors and video
motion to provide a preliminary liveness verification solution. Movee is
however severely limited, as (i) the ``verification'' step is constrained to
the initial few seconds of the video, and (ii) the system dictates the user to
pan the camera in a specific direction rather than gracefully accept the
natural motion of the user. These limitations significantly impact the
practical application of Movee.

Furthermore, Movee is vulnerable to the potent attacks that we study in this
paper. For example, an attacker starts Movee and points to a portion of a
target video playing on a projection screen, performs a pan motion as specified
by Movee, then points the camera to the whole frame of the fraudulent video.
Since Movee only uses the initial 6s chunk, the resulting sample passes Movee's
verifications.  Furthermore, in Section~\ref{sec:eval:clvamos} we
quantitatively show the ineffectiveness of Movee for free-form movements even
in 6s chunks: on the attacks we introduce, Movee's false positive rate is as
low as 38\% and its false negative rate is 28\%. 

We introduce Vamos to address these limitations and provide the first video
liveness verification system that works on unconstrained, free-form videos,
does not impose a ``verification'' step on users, and is resilient to a suite
of powerful, sensor based attacks.

This article extends our earlier work~\cite{RATC15} with the mirror attack and
two complex variants, the (i, p, c)-mirror and perturbed fingerprint attacks.
We show that Vamos performs surprisingly well on these attacks, and provide
insights into this result. We have also performed several new experiments:
(i) test Vamos on videos that belong to categories on which it was
not trained, (ii) test Vamos on instances fabricated according to attacks on
which it was not trained, and (iii) train Vamos on instances belonging to
all the video category and attack types, before testing it.

\newmaterial{
Wang et al.~\cite{WBRN15} proposed a video and sensor based human
fingerprinting technique. They use uploaded videos of an individual to extract
a motion fingerprint, which is a string of ``micro-activities'', e.g., walking
direction, stepping frequency, stopping, turning. Vamos does not fingerprint
humans in videos, but only needs to extract information about the camera
direction of movement.}

\newmaterial{Jain et al.~\cite{JMAB13} introduced
FOCUS, a Hadoop based video analytics system that uses 3D model reconstruction
and multimodal sensing to perform real time analysis and clustering of
user-uploaded video streams. While FOCUS is related to our video classification
effort, FOCUS is able to identify similar videos uploaded by multiple users,
e.g., of the same event, even if captured from radically different vantage
points.}

Several video watermarking algorithms has been proposed for video content
authentication~\cite{zhang2012video,chu2010error}. The goal of Vamos is however
not to authenticate the recorded video, but to verify the video liveness claim.
We note that watermarking only works if all the videos in the world employ it.
Furthermore, the defenses provided by invisible watermarks are defeated by
projection attacks.


Liu et al.~\cite{LLLS14} proposed a solution for summarizing (i.e., extracting
important frames from) mobile videos captured simultaneously with acceleration
and orientation streams. The acceleration values are used to exclude outliers.
Abdollahian et al.~\cite{ATPD10} define a ``camera view'' concept, and use
camera motion parameters to temporally segment, summarize and annotate user
generated videos. It will be interesting to evaluate a more efficient, video
summary based Vamos that identifies discrepancies between video and
acceleration summaries.

\section{System and Adversary Model}
\label{sec:model}

We consider a system that consists of a service provider (e.g.,
YouTube~\cite{YouTube}, iReport~\cite{iReport}, Stringwire~\cite{Stringwire})
and multiple subscribers. The service provider offers an interface for
subscribers to upload videos captured on their mobile devices.  We assume
subscribers own devices equipped with a camera and inertial sensors (i.e.,
acceleration).  Devices have Internet connectivity, which, for the purpose
of this work may be intermittent.  Each user is required to install an
application on her mobile device, which we henceforth denote as the ``client''.

The client simultaneously captures video and acceleration streams from the
device. It then uploads them to her account hosted by the service provider.
\newmaterial{ The provider uses Vamos to verify the data uploaded.
Specifically, Vamos divides the video and acceleration data into fixed length
chunks. For each chunk, it extracts motion vectors from the video and the
accelerometer streams, then generates features that encode the similarity of
the two motion vectors. Vamos uses these features to classify each chunk as
either genuine or fraudulent.  Vamos uses the results of the chunk level
classifications to extract a second set of features that model the similarity
of the original, whole video and acceleration data. It then uses supervised
learning to classify the uploaded video as genuine or fraudulent.} If genuine,
the provider makes the video publicly accessible, but keeps
the acceleration stream secret, or even discards it.


We assume that while the service provider is honest, users can be malicious.
Users can fraudulently claim ownership (creation) of videos they upload.  We
assume that attackers do not have access to the acceleration stream of videos
they intend to plagiarize. Thus, they are required to fabricate acceleration
streams for the uploaded videos.


%

We introduce several manual and automatic attacks that produce acceleration
streams that ``match'' targeted videos. \newmaterial{Vamos, or other sensor
based video liveness verification systems are not yet available in video
hosting sites. Thus, the attacks that we propose and implement here are
hypothetical.  However, our goal is to anticipate the adversarial strategy
changes likely to occur once video liveness verification systems are adopted.
This enables us to evaluate the resilience of Vamos to such strategy changes,
and be two steps ahead of adversaries.}

Let $\mathcal{A}$ denote the algorithm used by the attacker.  Let $V$ denote
the ``target'' video that the attacker wants to plagiarize.  Let
$\Gamma_{\mathcal{A}}$ denote additional information used by the attacker.  For
instance, $\Gamma_{\mathcal{A}}$ may include other videos and corresponding
acceleration streams. We denote the output of $\mathcal{A}$ as $\mathcal{A}(V,
\Gamma_{\mathcal{A}})$ = $\overline{Acc}$, the acceleration stream produced for
$V$.

We introduce first the ``sandwich'' attack, that enables the attacker to
manually produce the acceleration data. \newmaterial{This attack conjectures
that it should be easy for a human adversary to emulate the movement observed
in a target video and thus ``manually'' produce a matching accelerometer
stream.}

\noindent
{\bf Sandwich attack}.
The attacker studies the video $V$ and emulates the observed movement. For
instance, $\mathcal{A}$ stacks two devices. The attacker plays the target video
$V$ on the top device. He then moves the device stack to emulate the movement
seen on the top device.  The device on the bottom records the resulting
acceleration data, $\overline{Acc}$. $\mathcal{A}$ outputs $\overline{Acc}$.

We now describe the cluster attack: pair the target video with the acceleration
stream copied from a ``similar'' but genuine sample. \newmaterial{This attack
explores the hypothesis that an attacker with a large dataset of genuine
video and accelerometer streams, will be able to identify a pair whose
motion resembles the motion of the target video.}

\noindent
{\bf Cluster attack}.
$\mathcal{A}$ captures a dataset of genuine (video, acceleration) samples and
stores them in $\Gamma_{\mathcal{A}}$.  $\mathcal{A}$ uses a clustering
algorithm (e.g., K-means~\cite{Kmeans}) to cluster the videos based on their
movement. $\mathcal{A}$ classifies the target $V$ according to its movement and
assigns it to one of the previously generated clusters: the cluster containing
videos whose movement is closest to $V$.  $\mathcal{A}$ randomly picks one of
the genuine (video, acceleration) samples in the cluster. Let $(V', Acc')$ be
the chosen sample. $\mathcal{A}$ outputs $Acc'$.

\newmaterial{
We now introduce fully automatic ``mirror'' attacks, that copy the video motion
stream into the accelerometer stream. The hypothesis explored by this attack is
that since identical video and accelerometer streams will ``match'' perfectly,
they will be accepted as genuine.}

\noindent
{\bf Perfect mirror attack}.
Given a target video $V$, the adversary $\mathcal{A}$ extracts its video motion
stream \newmaterial{(e.g., using the VMA module of Section~\ref{sec:dam})} and
uses it to set the attack acceleration stream, $Acc$.  The output of the attack
consists of ($V$, $Acc$). Thus, this attack ``mirrors'' the video motion stream
into the acceleration stream.

\newmaterial{
In Section~\ref{sec:eval:clvamos} we show that the above hypothesis does not
hold: Vamos achieves perfect accuracy against the perfect mirror attack. This
is because a classifier easily learns that perfect matches are fraudulent.
In the following, we introduce two variations on this attack. The intuition
behind these two attacks is that perturbations in the accelerometer data
produce imperfect, but still ``close'' matches, resulting in genuine looking
video and accelerometer data.}

\noindent
{\bf (i, p, c)-mirror attack}.
After copying the motion stream extracted from the target video $V$, into the
fabricated acceleration stream $Acc$, the adversary $\mathcal{A}$ performs a
$(p, i, c)$ alteration of the stream.  Specifically, $\mathcal{A}$ inserts $i$
points in the accelerometer stream, randomly perturbs each accelerometer
reading between $-p$ and $p$ percent, then applies a calibration factor $c$. We
detail the process and the choice of the parameters, in
Section~\ref{sec:data:attack}.


\newmaterial{
The PFA attack fabricates an accelerometer stream as a patch of tiny snippets
that emulate real life data in the number of ``errors'' when matching against
the plagiarized video stream. We conjecture that such an accelerometer stream
produces a more realistic match. Specifically, PFA perturbs the copied video
motion stream for each snippet, to produce a video and acceleration snippet
that will have an amount of ``inconsistency'' that emulates the inconsistency
between a similar, but genuine video and acceleration snippet.}

\noindent
{\bf Perturbed fingerprint attack (PFA)}.
Given a target video $V$, PFA initializes the fraudulent acceleration
stream $Acc$ to the motion stream of $V$. PFA generalizes the (i, p, c)-mirror
attack. Unlike the (i, p, c)-mirror attack,
that uses a single set of perturbation factors, PFA splits $V$ and $Acc$ into
0.5s ``snippets'', and dynamically determines the amount of perturbation to be
applied to each snippet. For this, PFA leverages a ``dictionary'' that it
constructs from set of genuine video and acceleration samples. We provide
implementation details in Section~\ref{sec:data:attack}.


\newmaterial{
Next we introduce the stitch attack, that concatenates a plagiarized (video,
acceleration) chunk with several genuine chunks. The intuition
behind this attack is that video and accelerometer streams that differ only on
a few sections, will have a high similarity, due to the genuine chunks.  This
similarity will make it harder for a sensor based video liveness verification
solution to differentiate fraudulent and genuine data.} In
Section~\ref{sec:data:attack} we construct stitched samples from multiple
fraudulent and genuine chunks.

\noindent
{\bf Stitch attack}.
$\mathcal{A}$ takes as input parameters the target video $V$ and two integers,
$g > 0$ and $0 \le k \le g$. $\mathcal{A}$ first creates a set of genuine video
and acceleration chunks, $\Gamma_{\mathcal{A}}$ = $\{ (V_1, Acc_1), .., (V_1,
Acc_g) \}$, e.g., by capturing them on the mobile device.  $\mathcal{A}$ uses
one of the above attacks to fabricate $\overline{Acc}$, an acceleration stream
for $V$.  $\mathcal{A}$ then ``stitches'' the fake chunk $(V, \overline{Acc})$
with the genuine chunks $\Gamma_{\mathcal{A}}$, according to the index $k$. Let
$||$ denote the concatenation operation, applicable both to video and
acceleration streams. Then, $\mathcal{A}$ outputs $(V_a, Acc_a)$, where $V_a =
V_1 || ..  V_{k-1} || V || V_{k+1} .. || V_g$ and $Acc_a$ = $Acc_1$ $||$ ..
$Acc_{k-1}$ $||$ $\overline{Acc}$ $||$ $Acc_{k+1}$ .. $||$ $Acc_g$.

\noindent
\newmaterial{
{\bf Difficulty of implementing the attacks}.
Each of the above attacks require the attacker to write code, either to collect
sensor data, to implement a video motion based clustering algorithm, to copy
and perturb video motion data, or to stitch accelerometer streams from multiple
sources. We note that while an adversary can easily follow the attack
description to implement the code, this only needs to be done by one person,
who can then share or even sell it to regular, but adversarial users, who want
to thwart sensor based video liveness verification defenses.}

\newmaterial{
The cluster attack requires the adversary to manually collect a dataset of
video and accelerometer streams. While the dataset collection is a difficult
task, it is a one time effort. The sandwich attack, similarly requires manual
effort. However, different from the cluster attack, that effort needs to be
invested for each video targeted for plagiarism.}

\newmaterial{
We conclude that strictly automatic attacks (i.e., the mirror attack and
variants) are the easiest to perform. The manual attack (sandwich) and hybrid
manual and automatic attack (cluster) are harder to perform as they also
require human intervention. However, we note that the cluster attack can be
fully automated if attackers pool their resources to collect and share the
dataset of video and accelerometer streams. In addition, in our experiments,
see Section~\ref{sec:data:attack}, we observed that the two participants asked
to perform the sandwich attack on real videos, took only seconds to study the
target video and capture accelerometer data.}


\section{A Classification of Mobile Videos}
\label{sec:category}

\begin{table}[t]
\centering
\textsf{
\begin{tabularx}{0.45\textwidth}{XXXX} 
\hline
\textbf{Category ID} & \textbf{Distance to subject} & \textbf{User Motion} & \textbf{Camera Motion} \\
\hline\hline
1 & Close & Standing & Stationary \\ 
2 & Far & Standing & Stationary \\
3 & Close & Walking & Stationary \\
4 & Far & Walking & Stationary \\
\cmidrule[0.005em](lr{.75em}){1-4}
5 & Close & Standing & Scanning \\
6 & Far & Standing & Scanning \\
7 & Close & Walking & Scanning \\
8 & Far & Walking & Scanning \\
\cmidrule[0.005em](lr{.75em}){1-4}
9 & Close & Standing & Following \\
10 & Far & Standing & Following \\
11 & Close & Walking & Following \\
12 & Far & Walking & Following \\ [1ex] 
\hline 
\end{tabularx}
}
\vspace{5pt}
\caption{Video motion categories, based on (i) camera distance to the subject,
(ii) the user motion and (iii) camera motion.}
\vspace{-10pt}
\label{table:classification} 
\end{table}

\noindent
\newmaterial{
We posit that the success rate of the attacks previously introduced depends on
the type of motions encoded in the video. For instance, it seems intuitive that
videos where the hand-held device is stationary are easier to plagiarize.  To
verify our conjecture, we propose a general classification of videos captured
on mobile devices, based on the following dimensions:}

\begin{compactitem}

\item
{\bf User motion:}
We consider two types of recorder motions, ``standing'' and ``walking'', but no
motions such as jumping or driving.

\item
{\bf Camera motion:}
We consider three types of camera motions: ``stationary'', ``scanning'' and
``following''.  ``Scanning'' means the camera moves in a direction (e.g., left
to right) at a pace independent of the subject of the video. ``Following''
means the camera moves to maintain the subject within the confines of the
video. We have not considered videos shot with head mounted cameras.

\item
{\bf Distance to subject:}
We consider video subjects that are either ``close'' or ``far'' to the camera.
If the camera focuses on the subject of the video and only a limited area of
the background is observed in the video, we say the subject is ``close''.
Otherwise, the subject is ``far''.

\end{compactitem}

\noindent
Table \ref{table:classification} shows the resulting 12 mobile video
categories. Figure~\ref{fig:piechart} shows the category distribution of
YouTube and free-form video sets we collected (\S~\ref{sec:data:youtube}
and \S~\ref{sec:eval:youtube}).

\section{Vamos: Video Accreditation\\ Through Motion Signatures}
\label{sec:vamos}


\begin{figure}[!t]
\centering
\includegraphics[width=0.47\textwidth]{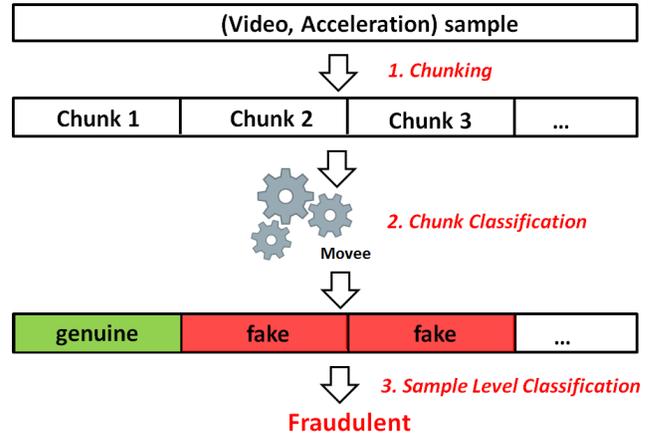}\\
\vspace{-10pt}
\caption{Illustration of the Vamos architecture and operation. Vamos consists
of three steps, (i) ``chunking'', to divide the (video, acceleration) sample,
(ii) chunk level classification, and (iii) sample level classification.}
\label{fig:vamos}
\vspace{-10pt}
\end{figure}

\noindent
In this section we introduce Vamos (Video Accreditation Through Motion
Signatures) an un-constrained video liveness analysis system. The
verifications of Vamos leverage the entire video and acceleration sample.  This
is in contrast with Movee, that relies only on the initial section of the
sample. Vamos consists of the three step process illustrated in
Figure~\ref{fig:vamos}.  First, it divides the input sample into equal length
chunks. Second, it classifies each chunk as either genuine or fraudulent.
Third, it combines the results of the second step with a suite of novel
features to produce a final decision for the original sample. In the following,
we detail each of these steps.

\vspace{-10pt}

\subsection{Chunk Extraction}

\noindent
The ``chunking'' process divides a video and acceleration sample $S = (V, Acc)$
into fixed length chunks. We consider a 1s granularity of division.  While 6s
is the chunk length we use in the experiments, we consider here a parameter $l$
to denote the length in seconds of the chunks. We call a \textit{transition
point} to be the time when the sample transitions from one video motion
category to another (e.g., from category 4 to category 8). Let a
\textit{transition chunk}, denote a $l$ second chunk that contains parts that
belong to multiple video categories.  Let $V[s,t]$ and $Acc[s,t]$ denote a
segment of $V$ and $Acc$ that starts at second $s$ and ends at second $t$.  The
chunking process produces a set $C$ of chunks, initially empty.  Let $L$ denote
the length of the $(V, Acc)$ sample.

\begin{figure}[!t]
\centering
\vspace{5pt}
\includegraphics[width=0.49\textwidth]{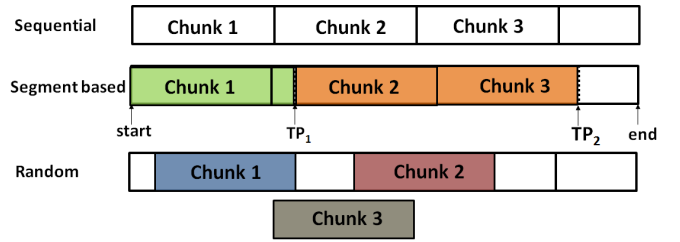}\\
\vspace{-5pt}
\caption{Chunk extraction illustration. For segment based chunking, the
first segment produces a single usable chunk. For random chunking, chunk 3
overlaps both chunks 1 and 2.}
\label{fig:chunking}
\vspace{-10pt}
\end{figure}

We propose three chunking techniques, illustrated in Figure~\ref{fig:chunking}.
\newmaterial{Sequential chunking is the obvious way of dividing arbitrary
length streams into chunks. Segment based chunking attempts to make the
division such that each chunk has a consistent video motion class. The output
of sequential and segment based chunking is easy to predict by an adversary.
Randomized chunking addresses this problem, by performing the division at
random positions.}

\noindent
{\bf Sequential chunking}.
Divide $(V, Acc)$ into sequential chunks, starting with the beginning. Let $n =
|C| = \lceil L/l \rceil$. Then, $C$ =$\{ (V[0,l-1], Acc[0,l-1])$, $(V[l,
2l-1], Acc[l, 2l-1]) .. (V[l (c-1), lc -1 ], Acc[l (c-1), lc - 1]) \}$.

\noindent
{\bf Segment based chunking}.
Identify the transition points of the sample $(V, Acc)$. Let a sample
\textit{segment} denote the part of a sample between either (i) the beginning
of the sample and the first transition point, (ii) two transition points, or
(iii) the last transition point and the end of the sample. Discard all segments
of $(V, Acc)$ whose length is less than $l$. Divide remaining segments
according to the sequential chunking described above.

\noindent
{\bf Randomized chunking}.
Produces $k$ chunks, $0 < k \le L$, where $k$ is an input argument, as follows.
Generate $k$ different index values
within the sample, $0 \le i_1,.., i_k \le L$ such that for any $s$ and $t$,
$1 \le s,t \le k$, $i_s + l \neq i_t$.
For each $i_j$, $j=1..k$, if $i_j \le L-l$, then $C = C \cup
(V[i_j, i_j+l-1], Acc[i_j, i_j+l-1])$.  Otherwise, $C = C \cup (V[i_j-l,
i_j], Acc[i_j-l, i_j])$.

Sequential chunking may produce transition chunks, that contain one or more
transition points. Segment based chunking will not produce transition chunks.
However, segment based chunking requires a mechanism to identify transition
points. Randomized chunking can produce transition chunks and also overlapping
chunks. Sequential and segment based chunking produce strictly non-overlapping
chunks.

\vspace{-5pt}

\subsection{CL-Vamos: Chunk Level Verification}
\label{sec:dam}

In the second step, Vamos classifies each chunk produced by the first step, as
either genuine or fraudulent. While Movee~\cite{RTC13} works on fixed length
chunks, it is limited to video and inertial sensor streams that encode one of 4
movements (up, down, to the left, or to the right). Specifically, 3 of the 14
features of Movee are (i) the placement of the bullseye relative to the center
of the screen, (ii) the dominant video motion direction and (iii) the dominant
sensor motion direction.

We introduce here CL-Vamos, the first liveness verification solution that works
on free form chunks, that encode unrestricted movements.  Similar to Movee,
CL-Vamos analyzes the consistency of the inferred motion from the
simultaneously and independently captured video and acceleration streams.
First, it uses an efficient image processing method to infer a motion vector
over the timeline of the video from frame-by-frame progress.  Second, it
converts the raw inertial sensor readings into a motion vector over the same
timeline. Third, CL-Vamos uses the Dynamic Time Warping algorithm
(DTW)~\cite{BookChapter4} to find the set of operations that minimizes the cost
of converting one vector to the other.
\newmaterial{We now detail each of these steps. The first two steps are
borrowed from Movee~\cite{RTC13} and are thus only briefly described.}

\noindent
\newmaterial{
{\bf Video Motion Analysis (VMA)}.
Given as input the captured video stream, the VMA module outputs an estimate
for the direction and magnitude of the camera movement. Given pairs of
consecutive frames, VMA uses ``phase correlation'' to find the movement of the
camera and output a frame-by-frame displacement vector. Specifically, phase
correlation finds the shift between two consecutive frames that minimizes the
difference between the two frames. It leverages the Fourier shift property: a
shift in the spatial domain of two images results in a linear phase difference
in the frequency domain of the Fourier Transform (FT)~\cite{Fourier}. It then
performs an element-wise multiplication of the transformed images, computes the
inverse Fourier transform (IFT) of the result, and finds the shift that
corresponds to the maximum amplitude - producing the resultant displacement.}

\newmaterial{
VMA applies the phase correlation method to obtain linear shifts between frames
in both the x and y axes. It then computes the \textit{cumulative shift} along
the x and y axes by adding up the linear shifts for all consecutive frames
retrieved from that video. The cumulative shifts on the x and y axes, along
with the orientation of the camera, enable VMA to infer the camera direction of
movement. We use the cumulative shifts and camera motion direction as feature
descriptors.}

\noindent
\newmaterial{
{\bf Inertial Sensor Motion Analysis (IMA)}.
The Inertial Sensor Motion Analysis (IMA) module leverages the accelerometer
sensor in order to produce a motion direction and magnitude; this data will be
used to verify its consistency with the output of the VMA module. Given the raw
accelerometer data, captured at a 16Hz frequency, the IMA module uses a
combination of low-pass and high-pass filters in order to remove the effects of
gravity. IMA then uses the filtered accelerometer data to extract the motion
direction and distance of the device, producing 3 cumulative shift vectors,
one along each of the x, y and z axes.}

\noindent
\newmaterial{
{\bf Similarity Computation (SC)}.
The Similarity Computation (SC) module compares the two motion sequences
computed by the VMA and the IMA modules. It returns a set of features that
summarize the nature of the similarity between the two sequences. In Movee, the
SC module computes the similarity of the two motion sequences on only one axis,
i.e., the dominant direction of movement.}

CL-Vamos is not restricted to the dominant direction of movement and removes
the features extracted from it. Instead, we have investigated a wide range of
features on both the x and y axes. Due to lack of space we report and evaluate
here (see Section~\ref{sec:eval}) the feature combination that achieved the
best performance.  Specifically, CL-Vamos computes the DTW between the motion
vectors extracted from the projections of the video and acceleration streams on
both the x and y axes. For each axis, DTW returns the number of diagonal,
expansion and contraction moves that convert one vector to the other, and the
cost of the resulting transformation. CL-Vamos uses this information to
generate the following features, for both the x and y axes:

\begin{compactitem}

\item
The DTW distance (transformation cost) between the video frame shift and
acceleration streams.

\item
The ratio of overlap points: the number of overlapping points in the two motion
vectors divided by the length of the vectors.

\item
The ratio of diagonal, expansion and contraction moves to the number of points
in the vectors.

\end{compactitem}

\noindent
{\bf Classification}.
\newmaterial{Manually identifying threshold values for the above metrics, that
would differentiate between fraudulent and genuine samples, is a difficult
task, made even more complex by the adversarial setup of the problem: the
adversary could exploit knowledge of explicit threshold values. Instead, we
leverage supervised learning algorithms to perform this task automatically,
learning threshold values from multiple genuine and fraudulent data samples.}

Specifically, CL-Vamos uses the above features, along with other Movee features
(e.g., the cumulative shift of the video and accelerometer on the $x$ and $y$
axes), with supervised learning to train classifiers. For each chunk $C_i$ in
$C$, let $c_i \in \{genuine, fake\}$ denote the classification produced by
CL-Vamos, and let $a_i \in \{genuine, fake\}$ denote the actual status of the
chunk.  We consider a ``positive'' to denote a fake chunk, and a ``negative''
to denote a genuine chunk.


We observe that the false positive rate of CL-Vamos, FPR = $Pr(c_i = fake |
a_i = genuine)$. That is, the false positive rate denotes the probability that
a chunk is classified as fake (positive), given that the chunk is in fact
genuine.  Similarly, the false negative rate is FNR = $Pr(c_i = genuine | a_i =
fake)$, the true positive rate is TPR = $Pr(c_i = fake | a_i = fake)$ and the
true positive rate is TNR = $Pr(c_i = genuine | a_i = genuine)$.\\

\vspace{-10pt}

\subsection{Vamos: Whole Video Classification}

\noindent
Let us assume that for a sample $S = (V, Acc)$, $f$ chunks in $C$ have been
classified as fraudulent and $g$ chunks have been classified as genuine.  Let
$n = f+g = |C|$.  We say $S$ is genuine iff $\forall i = 1 .. n$, $a_i$ =
``gen''. $S$ is fake if $\exists i$, $i = 1 .. n$, s.t., $a_i$ = ``fake''.
We can write the probability that the sample $S = (V, Acc)$ is fake,
$Pr(S = fake)$, given the above classification result, as 
\[
Pr[S = fake|\bigwedge_{i=1}^{g} (c_i = gen), \bigwedge_{i=g+1}^{n} (c_i = fake)] =
\]
\[
= 1 - \Pi_{i=1}^g Pr(a_i = gen| c_i = gen) \times
\]
\[
\Pi_{i=g+1}^n Pr(a_i =
gen| c_i = fake).
\]

\noindent
Let $\alpha = Pr(a_i = gen | c_i = gen)$, for any of the
chunks $C_i$ in $C$. Similarly, let $\beta = Pr(a_i = gen | c_i = fake)$.
Then, we have that
%
%
$Pr(S=fake) = 1 - \alpha^g \times \beta^f$.
Now, based on Bayes' theorem, we have that\\
%
%
$\alpha = \frac{TNR \times Pr(a_i = genuine)}{TNR \times Pr(a_i = genuine) +
FNR \times Pr(a_i = fake)}$.
Similarly, we have that
%
%
$\beta = \frac{FPR \times Pr(a_i = genuine)}{FPR \times Pr(a_i = genuine) + TPR
\times Pr(a_i = fake)}$.
We can compute thus $\alpha$ and $\beta$ as a function of $Pr(a_i = genuine)$
and $Pr(a_i = fake)$. We obtain these probability values statistically, based
on the performance of CL-Vamos on a large number of chunks. Specifically, $Pr(a_i
= fake)$ = $\frac{Nr.\ of\ fake\ chunks}{Total\ nr.\ of\ chunks}$ and $Pr(a_i =
genuine)$ = $\frac{Nr.\ of\ genuine\ chunks}{Total\ nr.\ of\ chunks}$, see
Section~\ref{sec:eval:vamos}.

We introduce several mechanisms to classify samples as genuine or fraudulent.
First, we propose a majority voting approach, where a sample $S = (V, Acc)$ is
classified as fraudulent if more than a threshold of the chunks of $S$ have
been classified by CL-Vamos as fraudulent: $\frac{f}{f+g} > thr$.  The
threshold $thr$ is a parameter that will be determined experimentally. Second,
we consider a probabilistic approach that labels a sample as fake if
$Pr(S = fake) = 1 - \alpha^g \times \beta^f$ is larger than a threshold value.
We experiment with threshold values in Section~\ref{sec:eval:vamos}.
Third, we propose a classifier based approach, that uses the following 
novel features:

\begin{compactitem}

\item
{\bf Results of CL-Vamos}:
The number of fraudulent chunks, $f$ and the number of genuine chunks $g$. The
classification results $c_i$, $\forall i=1..n$.  The probability that the
sample $S$ is fake, $Pr(S = fake)$.

\item
{\bf Aggregate features}:
For each of the 18 features of CL-Vamos, compute the minimum, maximum, average
and standard deviation of the feature's values over $c_i$, $\forall i=1..n$,
as new features.

\end{compactitem}

\noindent
Vamos uses these features with supervised learning to train classifiers for
samples of arbitrary length and motion types.

\section{Data Collection}
\label{sec:data}

\noindent
We have collected datasets of citizen journalism videos from YouTube and of
free-form (video, accelerometer) samples from real users. We have also created
datasets of fraudulent samples following the attacks introduced in
Section~\ref{sec:model}.  In the following we detail each dataset.

\vspace{-5pt}

\subsection{YouTube Video Collection}
\label{sec:data:youtube}

\noindent
We have collected 150 random citizen journalism videos from YouTube, in the
following manner. First, we have identified relevant topics using Wikipedia's
``Current\\ Events'' site~\cite{Wiki.Events}, BBC~\cite{BBC} and CNN~\cite{CNN}.
They include political events (e.g., Ukraine, Venezuela, Middle East), natural
disasters (e.g., earthquakes, tsunamis, meteorite landing), extreme sports and
wild life encounters. We have used keywords from such events to identify videos
in YouTube that have been captured by a regular person, using a mobile camera.
We have discarded videos shot by a professional cameraman or using a head
mounted camera. We collected the 150 videos from 139 users accounts.
We have made public this list of videos~\cite{youtube.videos}. The total length
of the 150 videos is 13,107 seconds. We analyze this dataset in
Section~\ref{sec:eval:youtube}.

\vspace{-5pt}

\subsection{Free-Form Video Collection}
\label{sec:data:freeform}

In previous work, Saini et al.~\cite{SVOC13} collected a dataset of 473 video
clips (along with accelerometer and compass readings) \textit{simultaneously}
recorded by users attending the same performance events. Cricri et
al.~\cite{CDCMG11,CDCMG14} exploit auxiliary sensor data to infer information
about the video content.  They show that multiple user records of a common
scene can be used to detect generic and specific events.

Instead, we performed a user study to simultaneously collect mobile videos shot
by users without any motion restrictions, and the associated accelerometer
readings.  We first briefly describe the implementation of Vamos, then detail
the free-form video collection process.

\noindent
{\bf Vamos implementation}.
We have implemented the Vamos client using Android, and a server component using
C++, R and PHP. We used the OpenCV (Open Source Computer Vision)
library~\cite{OpenCV} for the video motion analysis.
%
We used Nexus 4 smartphones (Android OS Jelly Bean version 4.2 with 1.5 GHz
CPU) to experiment with Vamos. Nexus 4 captures video at 30 fps (frames per
second) and samples accelerometer readings at a rate of
16.67Hz~\cite{SensorDelay}. The code is available on the project
website~\cite{Vamos}.

\noindent
{\bf Ethical considerations}.
We have used the Vamos application to collect video and acceleration
samples from real life users.  We have worked with the Institutional Review
Board (protocol number IRB-13-0582) at FIU to ensure an ethical interaction
with the users and collection of the data.

\begin{table}
\begin{center}
\small 
\textsf{
\begin{tabular}{ l l | r r}
\toprule
\textbf{Category} & \textbf{Chunk count} & \textbf{Category} & \textbf{Chunk count}\\
\midrule
1 & 26 & 8 & 28\\
2 & 50 & 9 & 26\\
3\&7 & 82 & 10 & 35\\
4 & 18 & 11 & 28\\
5 & 44 & 12 & 22\\
6 & 42 & & \\
%
\bottomrule
\end{tabular}
}
\end{center}
\caption{Number of chunks of the free-form dataset, per category.
Details in Section~\ref{sec:eval:youtube}.
}
\label{table:chunks}
\vspace{-10pt}
\end{table}

\noindent
{\bf Free-form data set}.
We have collected data from 16 users\footnote{11 are males and 5 females, aged
23-32, occupation including biology, fashion design, unemployed, and software,
civil and electrical engineering}. Each user was asked to use Vamos, following
the instructions shown on the screen: move the device in any direction to
capture videos. Each user contributed 10 free-form videos and accelerometer
data, producing a free-form dataset of 160 videos. We have manually annotated
the free-form dataset video samples according to the categories described in
Table~\ref{table:classification}.

We have divided each sample of the free-form dataset into 6s chunks, using
segment based chunking (see Section~\ref{sec:vamos}), producing a total of 401
genuine chunks. Table~\ref{table:chunks} shows the distribution of the chunks
into categories. We have made the free-form dataset publicly
available~\cite{Vamos}.


\vspace{-5pt}

\subsection{Attack Datasets}
\label{sec:data:attack}

\noindent
We have used the free-form dataset (see Section~\ref{sec:data:freeform}) to
create the following attack datasets.

\noindent
{\bf Sandwich attack dataset}.
Two skilled users have performed the sandwich attack on the 160 free-form video
dataset.  We have used the following procedure, for each whole video (not at
chunk level).  The attacker watches the target video an unlimited
number of times. The attacker stacks two phones. The attacker plays the target
video on the top device. The bottom device records the acceleration readings
during the session. The attacker can shoot any number of takes, until satisfied
with the result.


We combine the original video with the resulting attack acceleration sample to
produce a ``sandwich sample''. We used the segment based chunking method to
divide each sandwich sample into 6s (video,acceleration). Thus, each sandwich
chunk corresponds to one of the free-form chunks. The sandwich chunk dataset
contains thus 401 genuine and 401 fraudulent chunks.

\noindent
{\bf Cluster attack dataset}.
We ran K-means clustering~\cite{Kmeans} on the free-form chunk dataset, to
cluster the chunks according to their motion (see Cluster attack). We applied
the v-fold cross-validation algorithm~\cite{Arlot07modelselection} to determine
the optimal number of clusters in our dataset.  The outcome was $K$ = 6. The
cluster attack dataset consists of two subsets, of genuine and fraudulent
(video, acceleration) chunks. We used the free-form chunk dataset as the
genuine data. To create the fraudulent subset, for each genuine chunk, we
randomly chose another chunk from the same (motion) cluster. We then coupled
the video from the first chunk with the inertial sensor data of the randomly
selected chunk. Thus, the genuine and fraudulent subsets of the cluster attack
dataset each contain 401 chunks.

\noindent
{\bf Perfect mirror attack dataset}.
For each of the 401 genuine chunks, we have created a perfect mirror fraudulent
chunk: We copied the video from the genuine chunk, and set its accelerometer
stream to be the motion stream we extract from the video. Thus, the perfect
mirror attack dataset consists of 401 genuine and 401 fraudulent chunks.

\noindent
{\bf (i, p, c)-mirror attack dataset.}
For each genuine chunk, we have created a fraudulent chunk. Similar to the (i,
p, c)-mirror attack, we have initialized the acceleration stream of the
fraudulent chunk with the video motion stream. Then, we performed a 3 step
alteration of the acceleration stream as follows. First, stretch the
acceleration stream: between each pair of consecutive acceleration readings, we
insert $i \in_R \{2,3\}$ new points. The value of each inserted reading is set
to be equal to the mean of the previous and next readings in the stream. The
reason for the stretching step is that in Vamos, the ratio of the number of
acceleration readings to the video motion readings is around 2.5.


Second, ``perturb'' each resulting accelerometer reading $a$, to a value chosen
randomly in the interval $[a (1 - p), a (1 + p)]$. We have chosen $p = 0.1$ for
our experiments, as it is sufficient to prevent suspicious, perfect matches
between the video and acceleration streams, and also to prevent the creation of
random acceleration streams. Third, multiply each resulting acceleration value
by an empirically chosen factor $c \in_R [1,2]$.  This step ``calibrates'' the
acceleration stream to the observed range of 1 to 2 higher than the video
motion readings, of genuine samples.

\noindent
{\bf Perturbed fingerprint attack (PFA) attack dataset.}
%
PFA uses 10\% (40) of the chunks in the 401 genuine chunk dataset, to build the
dictionary%
%
%
The dictionary consists of 10 buckets, each corresponding to one of the
intervals $[0, 10), [10, 20), .., [90-100]$.  For each of the 40 chunks, PFA
does the following.  First, split the chunk (video and acceleration) into 0.5s
long snippets. Compute the DTW of the snippet and generate the DTW's percentage
of match moves out of all the moves.  Identify the bucket whose range contains
this value, and add the snippet to it.  The dictionary contains $480 = 40
\times 6 \times 0.5$ genuine snippets.

PFA randomly picks half of the remaining 361 (401-41) chunks to be genuine, and
half to generate fraudulent chunks. It uses the dictionary to extend the (i, p,
c)-mirror attack and create the fraudulent chunks.
%
%
For each fraudulent chunk, PFA performs the following steps. First, replace the
acceleration stream with the motion stream extracted from the video. Split the
resulting sample into 0.5s long snippets. For each snippet (identified as $snp$
in the following), use the video part to identify the nearest neighbor (in
terms of motion) snippet from the dictionary. Use the bucket of the nearest
neighbor to determine the amount of perturbation to be applied to the
acceleration of $snp$.  Specifically, pick a random value $x$ within the
interval of the bucket of the nearest neighbor. $x$ percent of the
accelerometer readings of $snp$ will remain unchanged, to contribute to $x$\%
match moves. Randomly pick $1-x$\% of the accelerometer readings of $snp$, and
apply an (i, p, c)-mirror transformation to them. The $i$ and $c$ values are
set as in the (i, p, c)-mirror transformation. The perturbation factor $p$ is a
parameter to this dataset, see Section~\ref{sec:eval:clvamos}.

\begin{figure}[!t]
\centering
\includegraphics[width=0.39\textwidth]{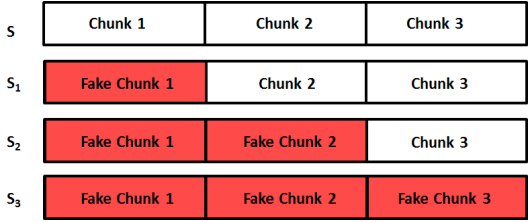}\\
\caption{Stitch attack example. For a genuine sample of 3 chunks, the attacker
produces 3 fake samples, with 1 to 3 fake (red) chunks. The genuine chunks are
copied from the genuine sample.}
\label{fig:stitch}
\vspace{-10pt}
\end{figure}

\noindent
{\bf Stitch attack datasets}.
We have built two stitch attack datasets, one based on the fake cluster chunks
and one on the fake sandwich chunks of the previous two attack datasets. The
construction process is the following. First, we discarded 4 out of the 160
free-form samples, as they do not have a 6s chunk belonging to a single
category.  We then discarded 43 samples that have only one chunk. For each of
the 113 remaining samples (that has at least 2 chunks), we construct 3
fraudulent samples.  For instance, for a 2 chunk genuine sample, we create a
fraudulent sample having the first chunk fake, the second genuine, one where
the first chunk is genuine, but the second is fake, and one where both chunks
are fake. For samples with more than 3 chunks, the position of the fake chunks
in any of the 3 created fake samples is randomly selected. The fake chunks are
from either the sandwich or the cluster chunk datasets.

Figure~\ref{fig:stitch} illustrates the generation of fraudulent samples given
a genuine free-form sample of 3 chunks. The reason for dropping samples with
less than 2 chunks is that we need to create the same number of fake samples
given any genuine sample (3 fakes per genuine sample). Samples with 1 chunk
cannot produce 3 fake stitch samples, thus had to be discarded.  The resulting
stitch datasets based on the cluster and sandwich attacks have thus each 339
fake samples ($(160-4-43) \times 3$).

\section{Evaluation}
\label{sec:eval}

We first report our experience in classifying the collected video datasets. We
then evaluate the ability of CL-Vamos to classify 6s chunks from the free-form
dataset, as either genuine or fraudulent. Finally, we evaluate the performance
of Vamos on the whole length samples from the free-form dataset.

\noindent
{\bf Metrics}.
In the following, the TPR (True Positive Rate) metric denotes the fraction of
videos correctly identified as fraudulent, the FPR (False Positive Rate)
denotes the fraction of videos incorrectly identified as fraudulent and the FNR
(False Negative Rate) denotes the fraction of videos incorrectly identified as
genuine. Accuracy denotes the ratio of the number of video correctly
classified (either as fraudulent or genuine) to the total number of videos
classified (including those correctly and incorrectly classified).

\vspace{-5pt}

\subsection{Video Dataset Classification}
\label{sec:eval:youtube}

\begin{figure}[!t]
\vspace{-10pt}
\centering
\includegraphics[width=0.39\textwidth,height=1.9in]{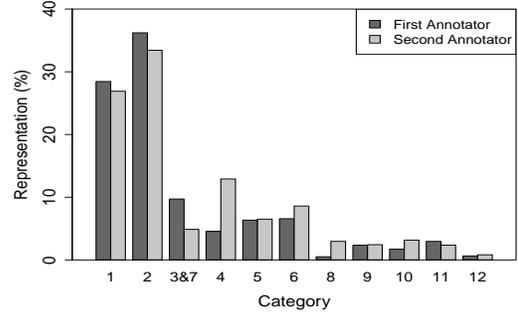}\\
\vspace{-5pt}
\caption{Distribution of motion categories of 150 YouTube citizen
journalism videos, produced by the manual annotation of two users. We observe a
consistent outcome of the annotation.}
\label{fig:youtubechart}
\vspace{-10pt}
\end{figure}

\noindent
Two users (paper authors) have manually annotated the YouTube and free-form
datasets based on the criteria described in Section~\ref{sec:category}. Since a
single video can include sections belonging to different motion categories, the
result of the annotation process consists of tuples of the form ($start\_time$,
$end\_time$, $category\_id$), where the first two fields denote the start and
end time of a video section (measured in seconds) and the last field denotes
the id of the video category (integer ranging from 1 to 12). At the end of the
process, we have computed a tally of the number of seconds of video belonging
to each of the 12 video motion categories.

Figure~\ref{fig:youtubechart} shows the percentage of each video motion
category in the collected videos, as produced by the two annotators.  We
observe a similar outcome for the two annotators. Motion categories 2 and 1
have the largest, whereas categories 12, 8, 9 and 10 have the smallest
representation.

The manual process enabled us to detect a discrepancy between the annotations
of two users, for categories 3 and 7.  Specifically, a walking user recording a
nearby scene without moving the camera, produces a video that can be (visually)
categorized as either 3 or 7. We have labeled these video with a separate
category denoted by ``3\&7''.  Similarly, when a standing user is following a
moving subject, but the subject is moving toward (or away from) the camera, the
camera movement of the resulting video can be interpreted as either following
or stationary.

\begin{figure}[!t]
\centering
\includegraphics[width=0.5\textwidth,height=1.75in]{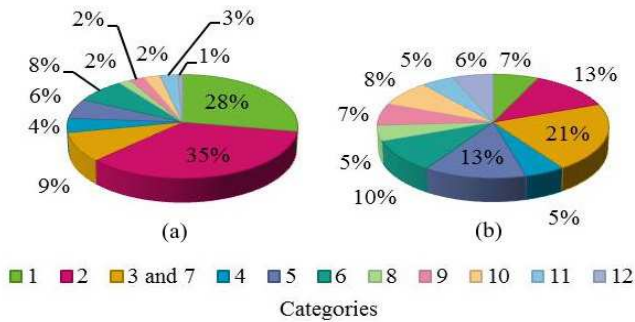}
\caption{(a) Motion category distribution for YouTube dataset.  (b)
Distribution for free-form dataset. Table~\ref{table:classification} defines
the 12 categories.
}
\label{fig:piechart}
\end{figure}

Figure~\ref{fig:piechart}(a) shows the overall category distribution of the
YouTube dataset, as an average of the distributions of the two annotators. The
motion categories 2 and 1 have the largest representation, whereas category 12
has the smallest representation.  Figure~\ref{fig:piechart}(b) shows the
category distribution of the free-form dataset, and Table~\ref{table:chunks}
shows the category distribution of the free-form chunks. The difference in
distributions of the YouTube and free-form datasets is likely due to the fact
that the free-form video collection scenarios have different dynamics from
citizen journalism scenarios.

\vspace{-5pt}

\subsection{Experiment Setup for CL-Vamos}
\label{sec:eval:setup}

CL-Vamos uses supervised learning algorithms to determine if a video is
genuine.  We have experimented with several classifiers, including MultiLayer
Perceptron (MLP)~\cite{MLP}, Decision Trees (DT) (C4.5), Random Forest
(RF)~\cite{RF2001} and Bagging~\cite{Breiman:1996}.  We have used the Weka data
mining suite~\cite{Weka} to perform the experiments, with default settings. For
the backpropagation algorithm of the MLP classifier, we set the learning rate
to 0.3 and the momentum rate to 0.2.

To compare CL-Vamos and Movee~\cite{RTC13} on the 11 video categories, we have
designed three experiments, that evaluate different scenarios.
In a ``category centric'' test, for each category $c$, we use
80\% randomly selected data from $c$ for training and the rest of 20\% for
testing.
%
%
This test evaluates the solution when only knowledge of the video's category
exists.
In a ``mixed data'' experiment, for each category $c$, we split the data into
10 equal sized folds.  Then, in each of 10 iterations, we train the classifier
on 9 folds from all the categories, and test on the data of the remaining fold
from $c$.  Repeat the procedure 10 times, ensuring each fold of $c$ appears
once in the test dataset.  This test evaluates performance when knowledge of
all the categories exists.
In the third, ``novelty'' test, for each category $c$, we train the supervised
learning algorithm on all the data from all the categories with the exception
of $c$. We then test the algorithm on all the data from $c$. This experiment
aims to predict performance on samples whose video motion category has not been
seen before.

\vspace{-5pt}

\subsection{CL-Vamos: Attack Detection}
\label{sec:eval:clvamos}

\begin{table}
\centering
\textsf{
\begin{tabular}{l c r r r r}
\toprule
\textbf{Category} & \textbf{TPR(\%)} & \textbf{FPR(\%)} & \textbf{FNR(\%)} & \textbf{Acc(\%)}\\
\midrule
1 & 83.33 & 75.0 & 16.67 & 60.0\\
2 & 63.63 & 44.44 & 36.36 & 60.0\\
3\&7 & 90.0 & 47.6 & 10.0 & 70.73\\
4 & 100.0 & 40.0 & 0.0 & 75.0\\
5 & 72.73 & 42.85 & 27.27 & 66.67\\
6 & 75.0 & 50.0 & 25.0 & 61.11\\
8 & 75.0 & 57.14 & 25.0 & 54.55\\
9 & 76.9 & 40.0 & 23.07 & 72.22\\
10 & 70.0 & 57.1 & 30.0 & 58.82\\
11 & 83.3 & 60.0 & 16.7 & 63.63\\
12 & 66.67 & 33.33 & 33.33 & 66.67\\
\bottomrule
\end{tabular}
}
\vspace{5pt}
\caption{CL-Vamos category centric experiment results for cluster
attack.  The accuracy is low, ranging between 54\% and 72\%.}
\vspace{-10pt}
\label{table:cluster:chunk:isolated}
\end{table}

\begin{table}
\centering
\textsf{
\begin{tabular}{l c r r r r}
\toprule
\textbf{Category} & \textbf{TPR(\%)} & \textbf{FPR(\%)} & \textbf{FNR(\%)} & \textbf{Acc(\%)}\\
\midrule
1 & 75.0 & 16.67 & 25.0 & 80.0\\
2 & 82.13 & 16.67 & 17.87 & 83.33\\
3\&7 & 87.97 & 27.0 & 12.03 & 77.08\\
4 & 75.0 & 25.0 & 25.0 & 75.0\\
5 & 80.0 & 2.86 & 20.0 & 83.33\\
6 & 68.33 & 13.9 & 31.67 & 79.17\\
8 & 75.0 & 0.0 & 25.0 & 80.0\\
9 & 77.66 & 16.67 & 22.34 & 80.0\\
10 & 91.67 & 38.86 & 8.33 & 75.0\\
11 & 83.25 & 6.25 & 16.75 & 85.0\\
12 & 75.0 & 25.0 & 25.0 & 81.25\\
\bottomrule
\end{tabular}
}
\vspace{5pt}
\caption{CL-Vamos mixed data experiment results for cluster attack. The
accuracy is as high as 85\% (on category 11.) We observe that knowledge
of other categories significantly improves the accuracy of CL-Vamos.}
\vspace{-10pt}
\label{table:cluster:chunk:mix}
\end{table}

\begin{table}
\centering
\textsf{
\begin{tabular}{l c r r r r}
\toprule
\textbf{Category} & \textbf{TPR(\%)} & \textbf{FPR(\%)} & \textbf{FNR(\%)} & \textbf{Acc(\%)}\\
\midrule
1 & 80.8 & 30.7 & 19.2 & 75.0\\
2 & 92.0 & 18.0 & 8.0 & 87.0\\
3\&7 & 96.1 & 21.6 & 3.9 & 87.25\\
4 & 81.0 & 4.8 & 19.0 & 88.09\\
5 & 93.3 & 26.7 & 6.7 & 83.33\\
6 & 81.8 & 27.3 & 18.2 & 77.27\\
8 & 89.3 & 21.4 & 10.7 & 83.93\\
9 & 84.4 & 42.2 & 15.6 & 71.11\\
10 & 93.0 & 27.9 & 7.0 & 82.56\\
11 & 92.9 & 32.1 & 7.1 & 80.36\\
12 & 95.5 & 18.2 & 4.5 & 88.64\\
\bottomrule
\end{tabular}
}
\vspace{5pt}
\caption{CL-Vamos novelty experiment results for cluster attack.
Knowledge of data from all other categories enables CL-Vamos to accurately
classify a novel category.}
\vspace{-10pt}
\label{table:cluster:chunk:novelty}
\end{table}

\noindent
{\bf Detection of the cluster attack}.
We first compare the per category performance of CL-Vamos and Movee  on the
cluster attack dataset (see Section~\ref{sec:data:attack}).  For CL-Vamos, we
have used the Random Forest algorithm, as it performed the best in our
experiments.  Table~\ref{table:cluster:chunk:isolated} shows our results for
the ``category'' experiment, including the true positive rate (TPR), false
positive rate (FPR), false negative rate (FNR) and accuracy (Acc).  We observe
a low accuracy, ranging between 54\% (category 8) and 72\% (category 9).
Table~\ref{table:cluster:chunk:mix} shows the TPR, FPR, FNR and accuracy
results for CL-Vamos on the ``mixed data'' experiment. The per category
accuracy ranges between 75\% and 85\%, a significant increase over the
``category'' experiment. This shows that additional data, even from other
categories, can benefit the performance of CL-Vamos.
This is confirmed by the ``novelty'' experiment, see
Table~\ref{table:cluster:chunk:novelty}: While its lowest accuracy is
71\%, its highest accuracy is 88\%.  Thus, even with knowledge
only from other categories, CL-Vamos can accurately classify chunks from
a newly seen category.

\begin{figure}
\centering
\vspace{-15pt}
\includegraphics[width=0.35\textwidth]{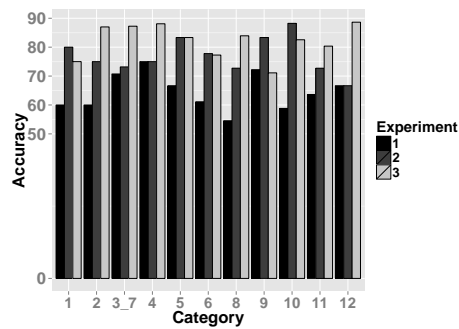}
\vspace{-5pt}
\caption{Summary of CL-Vamos accuracy on chunk-level cluster
attack dataset.  The novelty test results are surprisingly good: even without
knowledge of a category, CL-Vamos achieves an accuracy ranging between 71\% and
88\% (for categories 4 and 12 respectively).
\label{fig:accuracy:cluster:chunk}}
\vspace{-10pt}
\end{figure}

Figure~\ref{fig:accuracy:cluster:chunk} summarizes the per-category accuracy
achieved by CL-Vamos in the category centric, mixed data and novelty
experiments. In the mixed data and novelty experiments, CL-Vamos consistently
achieves better performance than on the category centric experiment.

\begin{table}
\centering
\textsf{
\begin{tabular}{l c r r r r}
\toprule
\textbf{Category} & \textbf{TPR(\%)} & \textbf{FPR(\%)} & \textbf{FNR(\%)} & \textbf{Acc(\%)}\\
\midrule
1 & 83.33 & 50.0 & 16.7 & 70.0\\
2 & 60.0 & 20.0 & 40.0 & 70.0\\
3\&7 & 77.78 & 30.0 & 22.22 & 73.68\\
4 & 66.67 & 0.0 & 33.33 & 87.5\\
5 & 66.67 & 33.33 & 33.33 & 66.67\\
6 & 80.0 & 33.33 & 20.0 & 75.0\\
8 & 80.0 & 40.0 & 20.0 & 70.0\\
9 & 60.0 & 28.57 & 40.0 & 64.71\\
10 & 70.0 & 16.7 & 30.0 & 75.0\\
11 & 75.0 & 42.9 & 25.0 & 63.63\\
12 & 100.0 & 50.0 & 0.0 & 87.5\\
\bottomrule
\end{tabular}
}
\vspace{5pt}
\caption{CL-Vamos category centric experiment results for the
sandwich attack dataset. The accuracy ranges between 64\% and 87\%, a marked
improvement over the corresponding experiment on the cluster attack dataset.}
\vspace{-10pt}
\label{table:sandwich:chunk:isolated}
\end{table}

\begin{table}
\vspace{-5pt}
\centering
\textsf{
\begin{tabular}{l c r r r r}
\toprule
\textbf{Category} & \textbf{TPR(\%)} & \textbf{FPR(\%)} & \textbf{FNR(\%)} & \textbf{Acc(\%)}\\
\midrule
1 & 75.0 & 10.0 & 25.0 & 84.0\\
2 & 66.67 & 16.67 & 33.33 & 73.33\\
3\&7 & 81.34 & 22.58 & 18.66 & 78.75\\
4 & 83.34 & 12.5 & 16.66 & 80.0\\
5 & 66.0 & 31.32 & 34.0 & 68.75\\
6 & 69.34 & 28.0 & 30.66 & 70.0\\
8 & 86.0 & 6.66 & 14.0 & 88.0\\
9 & 74.34 & 20.0 & 25.66 & 84.0\\
10 & 66.67 & 8.0 & 33.33 & 77.14\\
11 & 83.34 & 27.34 & 16.66 & 72.0\\
12 & 76.68 & 0.0 & 23.32 & 85.0\\
\bottomrule
\end{tabular}
}
\vspace{5pt}
\caption{CL-Vamos mixed data experiment on the sandwich attack dataset.
Accuracy ranges from 68\% to 88\%.}
\vspace{-10pt}
\label{table:sandwich:chunk:mix}
\end{table}

\begin{table}
\centering
\textsf{
\begin{tabular}{l c r r r r}
\toprule
\textbf{Category} & \textbf{TPR(\%)} & \textbf{FPR(\%)} & \textbf{FNR(\%)} & \textbf{Acc(\%)}\\
\midrule
1 & 80.8 & 34.6 & 19.2 & 73.08\\
2 & 68.0 & 22.9 & 32.0 & 72.45\\
3\&7 & 70.6 & 25.6 & 29.4 & 72.34\\
4 & 95.0 & 33.3 & 5.0 & 81.58\\
5 & 77.78 & 44.18 & 22.23 & 67.05\\
6 & 77.3 & 34.2 & 22.7 & 71.95\\
8 & 89.3 & 20.8 & 10.7 & 84.62\\
9 & 53.33 & 24.39 & 46.67 & 63.95\\
10 & 79.1 & 37.8 & 20.9 & 71.25\\
11 & 64.3 & 32.1 & 25.7 & 66.07\\
12 & 90.9 & 0.0 & 9.1 & 95.0\\
\bottomrule
\end{tabular}
}
\vspace{5pt}
\caption{CL-Vamos novelty experiment results for the sandwich attack
dataset. The accuracy ranges between 63\% (category 9) and
95\% (category 8). The improvement is thus not consistent across all the
video categories.}
\vspace{-5pt}
\label{table:sandwich:chunk:novelty}
\end{table}

\begin{figure}
\centering
\vspace{-15pt}
\includegraphics[width=0.35\textwidth]{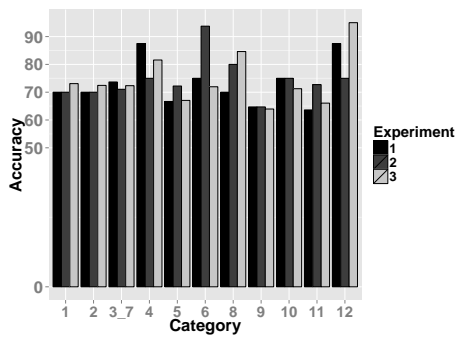}
\vspace{-5pt}
\caption{CL-Vamos accuracy on the sandwich attack
dataset, on the category centric, mixed data and novelty experiments. It
performs best on the mixed data experiment, but exhibits wide variability
across categories.
\label{fig:accuracy:sandwich:chunk}}
\vspace{-10pt}
\end{figure}

\noindent
{\bf Detection of the sandwich attack}.
We now evaluate CL-Vamos on the sandwich attack dataset of
Section~\ref{sec:data:attack}.
Table~\ref{table:sandwich:chunk:isolated} shows
the its performance on the category centric experiment. Its accuracy
ranges between 64\% and 87\%, showing a marked improvement over the
corresponding experiment on the cluster attack dataset.
Table~\ref{table:sandwich:chunk:mix} shows the per-category TPR, FPR, FNR and
accuracy results of CL-Vamos on the mixed data experiment. The results show an
improvement over both the category centric experiment, and over the mixed data
experiment on the cluster attack dataset.
Table~\ref{table:sandwich:chunk:novelty} shows the results of CL-Vamos on the
novelty experiment: its accuracy can be as high as 95\%.
Figure~\ref{fig:accuracy:sandwich:chunk} summarizes the per-category accuracy
achieved by CL-Vamos in the category centric, mixed data and novelty
experiments.


\noindent
{\bf Conclusion \#1}:
The results of the novelty experiment for the cluster and sandwich attack
show that CL-Vamos can be expected to perform quite well even if presented
with videos that belong to categories that might not have been considered
in our video motion classification.

\noindent
{\bf Category dependency}.
We now verify the intuition that the variation in FNR, FPR and accuracy of
CL-Vamos is due to its dependence on the video motion categories. We have
performed both Pearson's $\chi^2$ and Fisher's exact test on the results of
CL-Vamos for the sandwich attack dataset. The null hypothesis is that the true
positive, false positive, true negative and false negative values are
independent of the proposed video categories.
For the category centric test, the $\chi^2$'s p-value is $0.00044$ and Fisher's
p-value is $0.00013$. For the mixed data experiment, the $\chi^2$'s p-value is
$0.0001166$ and Fisher's p-value is $0.00015$. For the novelty
experiment, the $\chi^2$'s p-value is 0.01771 and Fisher's p-value is 0.00932.
Thus, we reject the null hypothesis and conclude that the performance of
CL-Vamos depends on the video motion category.

\noindent
{\bf Conclusion \#2}.
While we expected that certain motion categories are easier to plagiarize, our
results are surprising: CL-Vamos does not perform worst on categories 1 and 2,
captured by a standing user with a stationary camera. Based on observations
from our experiments, we believe that CL-Vamos exploits the ability of
accelerometers to capture the small, involuntary hand shakes that occur during
video capture sessions.  Instead, CL-Vamos has high FPR values for the sandwich
attack on categories 5, 6 and 11. This shows that in our experiments, humans
are better at plagiarizing videos shot while scanning or following subjects.

\noindent
{\bf Detection of the perfect mirror attack}.
CL-Vamos achieved 100\% accuracy when using either Random Forest, Decision Tree
or MLP on the perfect mirror attack dataset. The reason is that the following
features are consistently 0 for the fraudulent chunks: the DTW normalized
alignment distance, the penalized cost, the normalized number of moves
(expansion, contraction, match), and the normalized overlap. In addition, the
calibration factor is always 1. We conclude that the perfection of this attack
is also the reason for its failure.

\begin{table}
\centering
\textsf{
\begin{tabular}{l c r r r r}
\toprule
\textbf{Category} & \textbf{TPR(\%)} & \textbf{FPR(\%)} & \textbf{FNR(\%)} & \textbf{Acc(\%)}\\
\midrule
1 & 86.96 & 16.67 & 13.04 & 84.91\\
2 & 94.34 & 6.67 & 5.66 & 93.88\\
3 \& 7 & 94.0 & 6.93 & 6.0 & 93.53\\
4 & 73.33 & 4.54 & 26.66 & 86.48\\
5 & 96.65 & 4.76 & 4.35 & 95.45\\
6 & 100.0 & 0.0 & 0.0 & 100\\
8 & 73.68 & 0.0 & 26.32 & 88.37\\
9 & 92.6 & 0.0 & 7.4 & 97.02\\
10 & 90.91 & 15.79 & 9.09 & 87.32\\
11 & 100.0 & 0.0 & 0.0 & 100\\
12 & 77.78 & 22.22 & 22.22 & 80.0\\
\bottomrule
\end{tabular}
}
\vspace{5pt}
\caption{CL-Vamos mix experiment results for the (i, p, c)-mirror
attack.}
\vspace{-10pt}
\label{table:ipc}
\end{table}

\noindent
{\bf Detection of the (i, p, c)-mirror attack}.
We have run the mixed experiment to evaluate the performance of CL-Vamos on the
(i, p, c)-mirror attack datasets of Section~\ref{sec:data:attack}.
Table~\ref{table:ipc} shows the per-category performance of Vamos, that ranges
between 80\% and 100\%.

\begin{table}
\centering
\textsf{
\begin{tabular}{l c r r r r}
\toprule
\textbf{Algorithm} & \textbf{TPR(\%)} & \textbf{FPR(\%)} & \textbf{FNR(\%)} & \textbf{Acc(\%)}\\
\midrule
RF & 91.93 & 2.28 & 8.07 & 94.73\\
MLP & 90.86 & 2.86 & 9.14 & 93.91\\
DT & 89.78 & 2.86 & 10.22 & 93.35\\
\bottomrule
\end{tabular}
}
\vspace{5pt}
\caption{Overall performance of CL-Vamos on the PFA attack
dataset. Random Forest slightly outperforms MLP and Decision Tree. CL-Vamos 
exhibits high accuracy even on this complex attack.}
\vspace{-10pt}
\label{table:pfa}
\end{table}

\noindent
{\bf Detection of the PFA attack}.
We evaluated CL-Vamos against the PFA attack dataset in a manner different from
the other attacks. Specifically, we have run 10 different experiments. In each
experiment we have built a new PFA attack dataset: pick a new random dictionary
membership, and a different set of 361 genuine and fraudulent chunks. Then, we
ran 10-fold cross validation over each dataset. Table~\ref{table:pfa} shows the
average results over the 10 experiments. While the Random Forest classifier
performs best, all the classifiers tested with CL-Vamos achieved an accuracy
exceeding 93\%.


\noindent
{\bf Conclusion \#3}.
We have expected that the mirror attack and variants will be very effective
against CL-Vamos. Instead, we discovered that a perfect match can be used by
CL-Vamos to trivially detect fraud.  Furthermore, CL-Vamos is effective even
against mirror attack variants that introduce controlled perturbation to the
match. Based on an analysis of the features produced by CL-Vamos, we conjecture
that an observed accelerometer ``inertia'' is responsible for our success.
Specifically, we observed that when the user stops moving the device, the video
motion stream records this event, while the accelerometer continues to record
unusually high readings. Similarly, when the device begins to move, unlike the
video, the accelerometer experiences a delay in recording the motion. The
difficulty to emulate this inertia, which we conjecture occurs throughout the
motion of the device, enables the DTW features to effectively capture
differences between genuine and fraudulent samples.

\begin{table}
\centering
\textsf{
\begin{tabular}{l c r r r r}
\toprule
\textbf{Attack} & \textbf{TPR(\%)} & \textbf{FPR(\%)} & \textbf{FNR(\%)} & \textbf{Acc(\%)}\\
\midrule
Cluster & 100 & 19.51 & 0 & 90.12\\
Sandwich & 65.0 & 19.51 & 35.0 & 72.84\\
(p, i, c)-mirror & 97.62 & 19.51 & 2.38 & 89.16\\
PFA & 95.0 & 19.51 & 5.0 & 87.65\\
\bottomrule
\end{tabular}
}
\vspace{5pt}
\caption{Overall performance of CL-Vamos on mix attack
scenario, where CL-Vamos is trained on 90\% of all attack instances.
With the exception of the sandwich attack, the knowledge of other
attack instances greatly improves the TPR of CL-Vamos.}
\vspace{-10pt}
\label{table:mix:attack}
\end{table}

\noindent
{\bf Mixed attack evaluation}.
We have evaluated a scenario where CL-Vamos has knowledge of all the attacks.
Specifically, we have used all the attack datasets, for a total of 401 cluster,
401 sandwich, 401 (i, p, c)-mirror and 10 $\times$ 180 PFA attack instances.
We have trained CL-Vamos on 90\% of each of the attack datasets, as well as
90\% of the 401 genuine instances. We have then tested CL-Vamos separately on
the remaining 10\% of attack instance of each attack, along with 90\% of the
genuine instances.  To compensate for the 7 hold imbalance in the number of
fraudulent vs. genuine instances, we have assigned a penalty during training
for incorrectly classifying fraudulent (1/8)) and genuine instances (7/8).

Table~\ref{table:mix:attack} shows that with the exception of the sandwich
attack, the knowledge of other attack instances greatly improves the TPR of
CL-Vamos. The FPR increase is due to the imbalance between the number of
fraudulent and genuine instances used during training.

\begin{table}
\centering
\textsf{
\begin{tabular}{l c r r r r}
\toprule
\textbf{Attack} & \textbf{TPR(\%)} & \textbf{FPR(\%)} & \textbf{FNR(\%)} & \textbf{Acc(\%)}\\
\midrule
Cluster & 62.06 & 43.75 & 37.94 & 61.04\\
Sandwich & 22.68 & 8.75 & 77.32 & 33.88\\
(i, p, c)-mirror & 90.78 & 40.0 & 9.22 & 85.88\\
PFA & 93.33 & 42.5 & 6.67 & 83.45\\
\bottomrule
\end{tabular}
}
\vspace{5pt}
\caption{Overall performance of CL-Vamos on new attacks: CL-Vamos is tested on
data from attacks on which it was not trained.  The sandwich attack is
effective: Without knowledge of sandwich attack instances, CL-Vamos performs
poorly.}
\vspace{-10pt}
\label{table:new:attack}
\end{table}

\noindent
{\bf Resilience to new attacks}.
To investigate the ability of CL-Vamos to detect new, previously unseen
attacks, we have performed the following experiment. For each attack type $a$
we considered (cluster, sandwich, mirror variants), we trained CL-Vamos using
all the fraudulent instances from all the attack datasets except $a$, as well
as 80\% of all the genuine chunks. We have then tested CL-Vamos on all the
instances of the attack dataset of type $a$, and the remaining 20\% genuine
chunks.

Table~\ref{table:new:attack} shows the results.  As expected, the performance
of CL-Vamos degrades on all the attacks. In particular, for the sandwich
attack, CL-Vamos achieves an overall accuracy of 33.88\%. This result confirms
the strength of the sandwich attack: in the absence of any sandwich attack
training instances, CL-Vamos incorrectly accepts as genuine, more than 3 out of
4 fraudulent, sandwich chunks.

\vspace{-5pt}

\subsection{CL-Vamos vs. Movee}

\begin{figure}
\centering
\vspace{-20pt}
\includegraphics[width=0.39\textwidth]{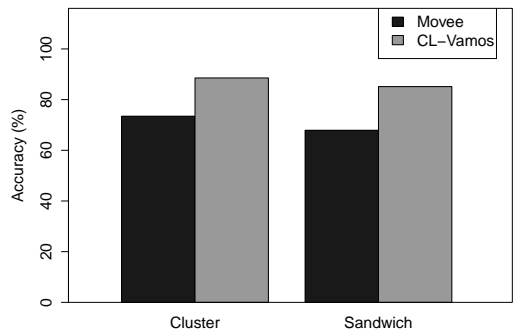}
\vspace{-25pt}
\caption{Accuracy of CL-Vamos and Movee on mixed data experiment. CL-Vamos
outperforms Movee by 15\%+, for both cluster and sandwich attacks.
\label{fig:vamos.movee}}
\vspace{-20pt}
\end{figure}

We have compared the performance of CL-Vamos and Movee using the mixed data
experiment, on the cluster and sandwich attack datasets.
Figure~\ref{fig:vamos.movee} summarizes our results. On the cluster attack,
CL-Vamos achieves 88\% accuracy when using MLP (Random Forest 83\%, Bagging
81\%, Decision Trees 78\% and SVM 80\%). Movee achieves highest accuracy when
using the Random Forest (73\%). On the sandwich attack, CL-Vamos achieves 85\%
accuracy when using Random Forest (MLP 71\%, Bagging 78\%, Decision Trees 74\%
and SVM 80\%).  Movee achieves the highest, 67\% accuracy, when using either
Random Forest or Bagging classifiers. This substantial improvement of CL-Vamos
corresponds to an FNR of 6-14\% and FPR of 15-17\% on these attacks. In
contrast, Movee's FNR is between 21-28\% and FPR is between 31-38\%.

\vspace{-5pt}

\subsection{CL-Vamos on Citizen Journalism}

Current YouTube videos do not have acceleration data. CL-Vamos however only
works for video chunks for which we have acceleration data. We propose to use
the classification of the collected YouTube videos (see
Section~\ref{sec:eval:youtube}) and the performance of CL-Vamos on the
free-form video and acceleration samples (see Section~\ref{sec:eval:clvamos}) to
predict its performance on fixed length chunks of citizen journalism videos
from YouTube.

Let $Acc(i,FreeForm,AT)$ denote the accuracy of CL-Vamos on videos from the
$i$-th category ($i$=1..11) of the free-form dataset, for a given attack type
AT.  We define the predicted accuracy of CL-Vamos for YouTube and the attack type
AT, $Acc_p(YouTube, AT)$, as the weighted sum of its per-category accuracy on
the free-form dataset:\\
$Acc_p(YouTube, AT)$ = $\sum_{i=1}^{11} w_i \times Acc(i,FreeForm,AT)$.
We define the weight $w_i$ to be the percentage of chunks of category $i$ in
the YouTube dataset, as shown in Figure~\ref{fig:piechart}(a). In the
YouTube dataset categories 1 and 2 have the highest weight. The predicted
accuracy of CL-Vamos for the cluster attack on the YouTube dataset is then
80.9\%, and for the sandwich attack is 77.19\%.





\vspace{-5pt}

\subsection{Vamos Evaluation}
\label{sec:eval:vamos}

We now evaluate the performance of Vamos on entire video and acceleration
samples. We note that a sample can consist of multiple chunks that belong to
different motion categories.  We have performed experiments using the stitch
attack datasets (based on chunk-level cluster and sandwich attacks) described
in Section~\ref{sec:data:attack}. The stitch attack datasets consist of both
genuine and fraudulent free-form samples.

\begin{figure}
\centering
\includegraphics[width=0.49\textwidth]{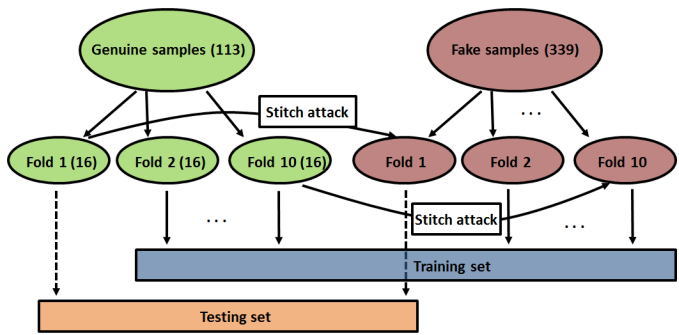}
\vspace{-5pt}
\caption{Setup of Vamos experiment. Each genuine fold
produces a stitch attack fold. In each experiment, 9 genuine
folds and the corresponding stitch attack folds are used for training. The rest
are used for testing.
\label{fig:vamos:experiment}}
\vspace{-10pt}
\end{figure}

Vamos makes the sample level classification decision based on the
classification of the chunks of the sample. In order to avoid a case where the
same chunk appears in both training and testing sets, we propose the following
experimental design, illustrated in Figure~\ref{fig:vamos:experiment}.

First, divide the dataset of 113 samples of at least 2 chunks each, into $k$
folds, $gen.fold(i)$, $i=1..k$, and the corresponding 339 attack sample dataset
(either cluster or sandwich attack based) into $k$ folds, $attack.fold(i)$,
$i=1..k$. The split takes place such that the samples from the $gen.fold(i)$
were used to generate the attack samples from $attack.fold(i)$. Then, for each
$i=1..k$, pick all the samples from $gen.fold(j)$ and $attack.fold(j)$,
$j=1..k$, $j\neq i$, and use their chunks to train CL-Vamos. Run the trained
CL-Vamos on all the chunks from $gen.fold(i)$ and $attack.fold(i)$. Given the
classified chunks of the samples from $gen.fold(i)$ and from $attack.fold(i)$, run
the sample level classification step to classify the samples. For instance, to
compute $Pr(S=fake)$ for a sample $S$, compute $Pr(a_i=genuine)$ and
$Pr(a_i=fake)$ based on the number of fake and genuine chunks in the training
folds (see Section~\ref{sec:vamos}). In our experiments, we set $k$ to 10.

\begin{table}
\vspace{10pt}
\centering
\textsf{
\begin{tabular}{l r r r r r}
\toprule
\textbf{Algo} & \textbf{Thr} &  \textbf{TPR(\%)} & \textbf{FPR(\%)} & \textbf{FNR(\%)} & \textbf{Acc(\%)}\\
\midrule
%
%
\multirow{4}{0pt}{Maj. Vote} & 0.1 & 91.69 & 7.95 & 8.31 & 91.78\\
& 0.3 & 84.09 & 4.39 & 15.91 & 86.97\\
& 0.5 & 34.80 & 0.00 & 65.20 & 51.10\\
& 0.7 & 24.49 & 0.00 & 75.51 & 43.37\\ \hline \\[-5pt]
%
%
\multirow{3}{0pt}{Prob} & 0.6 & 92.55 & 26.21 & 7.45 & 87.86\\
& 0.7 & 91.69 & 7.95 & 8.31 & 91.78\\
& 0.8 & 61.39 & 5.38 & 38.61 & 69.70\\ \hline \\[-5pt]
Bagging & & 97.35 & 5.08 & 2.65 & 95.53\\
\bottomrule
\end{tabular}
}
\vspace{5pt}
\caption{Vamos efficacy on cluster based stitch attack.  The supervised
learning approach performs best.}
\vspace{-10pt}
\label{table:cluster:vamos}
\end{table}

\begin{table}
\centering
\textsf{
\begin{tabular}{l r r r r r}
\toprule
\textbf{Algo} & \textbf{Thr} & \textbf{TPR(\%)} & \textbf{FPR(\%)} & \textbf{FNR(\%)} & \textbf{Acc(\%)}\\
\midrule
%
\multirow{4}{0pt}{Maj. Vote} & 0.1 & 85.10 & 41.29 & 14.90 & 78.50\\
& 0.3 & 76.23 & 40.76 & 23.77 & 71.97\\
& 0.5 & 34.93 & 13.56 &  65.07 & 47.76\\
& 0.7 & 25.55 & 8.11 & 74.45 & 42.08\\ \hline \\[-5pt]
%
%
\multirow{3}{0pt}{Prob} & 0.6 & 93.76 & 75.53 & 6.24 & 76.44\\
& 0.7 & 91.67 & 70.08 & 8.33 & 74.64\\
& 0.8 & 66.46 & 34.92 & 33.54 & 66.12\\ \hline \\[-5pt]
Bagging & & 83.7 & 3.63 & 16.3 & 93.199\\
\bottomrule
\end{tabular}
}
\vspace{5pt}
\caption{Vamos performance on sandwich/stitch attack.  Neither of the majority
voting and the probabilistic approaches simultaneously achieves small FPR and
FNR values. The supervised learning approach significantly reduces both the FPR
and FNR values.} 
\vspace{-10pt}
\label{table:sandwich:vamos}
\end{table}

\noindent
{\bf Experiment results}.
Table~\ref{table:cluster:vamos} reports the performance of Vamos on the cluster
based stitch attack dataset. We have experimented with multiple threshold
values. For majority voting, a threshold of 0.1 performed best: both FPR and
FNR values are under 9\%.  For the probabilistic approach, a threshold of 0.7
achieves similar performance. However, we note that the classifier
approach, when using the Bagging algorithm, significantly outperforms the other
solutions, with an FPR of around 5\% and an FNR of 2.65\%.

\begin{figure}
\centering
\vspace{-25pt}
\includegraphics[width=0.39\textwidth]{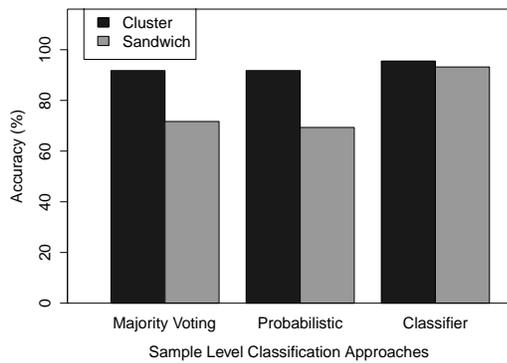}
\vspace{-15pt}
\caption{Vamos accuracy on stitch attacks.  Even for the sandwich stitch
attack, the classifier approach (using Bagging) exceeds 93\% accuracy.
\label{fig:vamos:accuracy}}
\vspace{-10pt}
\end{figure}

Table~\ref{table:sandwich:vamos} shows the performance of the majority voting,
probabilistic and classifier based approaches of Vamos, on the sandwich based
stitch attack dataset. For majority voting and probabilistic approaches, the
sandwich based stitch attack is significantly more efficient: The majority
voting has no threshold where both FPR and FNR are below 35\%. The
probabilistic approach achieves its optimum for a threshold of 0.8, when its
FPR and FNR values are barely under 35\%. In contrast, the classifier approach,
again when using Bagging, exhibits a significantly improved performance, with
an FPR of under 4\% and an FNR of 16.3\%.
%
%
Figure~\ref{fig:vamos:accuracy} summarizes the accuracy of the three approaches of
Vamos for the cluster and sandwich based stitching attacks.



%
%

\vspace{-5pt}

\section{Limitations}

\newmaterial{
Vamos needs to have access to inertial sensor data captured simultaneously with
the video. The clients of most live streaming apps do not currently upload this
data. Vamos can however be easily integrated with other solutions, e.g.,
FOCUS~\cite{JMAB13}, that use simultaneously captured sensor data to
automatically cluster mobile videos uploaded by users (see
Section~\ref{sec:related}).}

\newmaterial{
Vamos could benefit from the use of additional sensors, e.g., the gyroscope.
The ability to verify the consistency of the motion extracted from video,
accelerometer and gyroscope reading streams would likely make it more difficult
for an attacker to successfully launch the attacks that we proposed.  We note
however that Vamos achieves good accuracy even when leveraging only the
accelerometer.}

%

\vspace{-5pt}

\section{Conclusions}
\label{sec:conclusions}

We proposed Vamos, the first length and motion un-constrained video liveness
verification system. Vamos uses the entire video and acceleration streams to
identify video fraud. We proposed a suite of novel, manual and automatic
attacks that target acceleration based video liveness analysis solutions.  Our
evaluation, on data collected from a user study and on citizen journalism
videos from YouTube, shows that Vamos accurately detects even complex attacks.
In addition, we have introduced a motion based classification of videos, and
have shown that the accuracy of Vamos depends on the video category.


\vspace{-5pt}

\section{Acknowledgments}

This research was supported in part by NSF grant CNS-1526494 and
DoD grant W911NF-13-1-0142.

\vspace{-5pt}

\bibliographystyle{abbrv}

\begin{thebibliography}{10}

\bibitem{BBC}
{British Broadcasting Corporation}.
\newblock \url{http://www.bbc.com/}.

\bibitem{CNN}
{Cable News Network}.
\newblock \url{www.cnn.com}.

\bibitem{CitizenEvidenceLab}
{Citizen Evidence Lab}.
\newblock \url{http://citizenevidence.org/}.

\bibitem{CitizenTube}
Citizentube.
\newblock \url{http://www.citizentube.com/}.

\bibitem{iReport}
{CNN iReport. Explore. Discover. Contribute}.
\newblock \url{http://ireport.cnn.com/}.

\bibitem{InformaCam}
{InformaCam: Verified Mobile Media}.
\newblock \url{https://guardianproject.info/apps/informacam/}.

\bibitem{OpenCV}
{Open Source Computer Vision}.
\newblock \url{http://opencv.org/}.

\bibitem{SensorDelay}
{Sensor Delay}.
\newblock
  \url{http://developer.android.com/reference/android/hardware/SensorManager.html}.

\bibitem{Stringwire}
{Stringwire. Live Video. Made Social.}
\newblock \url{https://stringwire.com/}.

\bibitem{Vamos}
Vamos.
\newblock \url{http://users.cis.fiu.edu/~carbunar/caspr.lab/liveness.html}.

\bibitem{Weka}
{Weka}.
\newblock \url{http://www.cs.waikato.ac.nz/ml/weka/}.

\bibitem{Wiki.Events}
Wikipedia current events.
\newblock \url{http://en.wikipedia.org/wiki/Category:Current_events}.

\bibitem{Witness}
{Witness. See it. Film it. Change it.}
\newblock \url{witness.org}.

\bibitem{WitnessOrg}
{Witness.org}.
\newblock \url{http://witness.org/}.

\bibitem{YouTube}
{YouTube}.
\newblock \url{http://www.youtube.com}.

\bibitem{youtube.videos}
Youtube videos.
\newblock \url{http://seclab.cs.fiu.edu/resources}.

\bibitem{Guardian}
Us intelligence officials working to establish authenticity of video of sotloff
  being killed, reportedly by the same fighter who murdered james foley.
\newblock
  http://www.theguardian.com/world/2014/sep/02/isis-video-steven-sotloff-beheading,
  September 2014.

\bibitem{Fake.Germanwings}
{Fake Germanwings pictures circulate online}.
\newblock BBC Trending, \url{http://www.bbc.com/news/blogs-trending-32037008},
  March 2015.

\bibitem{Fake.MH370}
{The fake 'MH370 search' video that went viral}.
\newblock BBC Trending, \url{http://www.bbc.com/news/blogs-trending-26770110},
  March 2015.

\bibitem{Rohingya}
{The fake pictures of the Rohingya crisis}.
\newblock BBC Trending, \url{http://www.bbc.com/news/blogs-trending-32979147},
  June 2015.

\bibitem{Fake.Syrian.Hero}
{Video of ‘Hero Syrian Boy’ Saving Sister From Gunfire Was Fake —
  Here’s the Proof}.
\newblock The Blaze,
  \url{http://www.theblaze.com/stories/2014/11/15/video-of-hero-syrian-boy-saving-sister-from-gunfire-was-fake-heres-the-proof/},
  November 2015.

\bibitem{Android.FakeGPS}
{Fake gps - fake location}.
\newblock Google Play,
  \url{https://play.google.com/store/apps/details?id=com.fakegps.mock&hl=en},
  Last accessed in May 2015.

\bibitem{Android.FakeGPS1}
{Fake GPS location}.
\newblock Google Play,
  \url{https://play.google.com/store/apps/details?id=com.lexa.fakegps&hl=en},
  Last accessed in May 2015.

\bibitem{Android.FakeGPS2}
{Fake Location Spoofer Free}.
\newblock Google Play,
  \url{https://play.google.com/store/apps/details?id=com.incorporateapps.fakegps.fre&hl=en},
  Last accessed in May 2015.

\bibitem{ATPD10}
G.~Abdollahian, C.~M. Taskiran, Z.~Pizlo, and E.~J. Delp.
\newblock Camera motion-based analysis of user generated video.
\newblock {\em {IEEE} Transactions on Multimedia}, 12(1):28--41, 2010.

\bibitem{Arlot07modelselection}
S.~Arlot.
\newblock Model selection by resampling penalization, 2007.

\bibitem{Kmeans}
C.~M. Bishop.
\newblock {\em Neural Networks for Pattern Recognition}.
\newblock Oxford University Press, Inc., New York, NY, USA, 1995.

\bibitem{Breiman:1996}
L.~Breiman.
\newblock Bagging predictors.
\newblock {\em Mach. Learn.}, 24(2):123--140, Aug. 1996.

\bibitem{RF2001}
L.~Breiman.
\newblock Random forests.
\newblock {\em Machine Learning}, 45:5--32, 2001.

\bibitem{chu2010error}
S.-C. Chu, L.~C. Jain, H.-C. Huang, and J.-S. Pan.
\newblock Error-resilient triple-watermarking with multiple description coding.
\newblock {\em Journal of Networks}, 5(3):267--274, 2010.

\bibitem{CDCMG11}
F.~Cricri, K.~Dabov, I.~Curcio, S.~Mate, and M.~Gabbouj.
\newblock Multimodal event detection in user generated videos.
\newblock In {\em Multimedia (ISM), 2011 IEEE International Symposium on},
  pages 263--270, Dec 2011.

\bibitem{CDCMG14}
F.~Cricri, K.~Dabov, I.~D. Curcio, S.~Mate, and M.~Gabbouj.
\newblock Multimodal extraction of events and of information about the
  recording activity in user generated videos.
\newblock {\em Multimedia Tools Appl.}, 70(1):119--158, May 2014.

\bibitem{Fourier}
J.~B.~J. Fourier and A.~Freeman.
\newblock {\em The Analytical Theory of Heat}.
\newblock Cambridge University Press, 2009.

\bibitem{MLP}
S.~I. Gallant.
\newblock Perceptron-based learning algorithms.
\newblock {\em Trans. Neur. Netw.}, 1(2):179--191, June 1990.

\bibitem{JMAB13}
P.~Jain, J.~Manweiler, A.~Acharya, and K.~Beaty.
\newblock Focus: Clustering crowdsourced videos by line-of-sight.
\newblock In {\em Proceedings of the 11th ACM Conference on Embedded Networked
  Sensor Systems}, 2013.

\bibitem{LLLS14}
Y.~Liu, H.~Liu, Y.~Liu, and F.~Sun.
\newblock User-generated-video summarization using sparse modelling.
\newblock In {\em Neural Networks (IJCNN), 2014 International Joint Conference
  on}, pages 3909--3915, July 2014.

\bibitem{BookChapter4}
M.~Müller.
\newblock Dynamic time warping.
\newblock In {\em Information Retrieval for Music and Motion}, pages 69--84.
  Springer Berlin Heidelberg, 2007.

\bibitem{RATC15}
M.~Rahman, M.~Azimpourkivi, U.~Topkara, and B.~Carbunar.
\newblock Liveness verifications for citizen journalism videos.
\newblock In {\em Proceedings of the 8th ACM Conference on Security \& Privacy
  in Wireless and Mobile Networks}, pages 17:1--17:10, 2015.

\bibitem{RTC13}
M.~Rahman, U.~Topkara, and B.~Carbunar.
\newblock {Seeing is Not Believing: Visual Veriﬁcations Through Liveness
  Analysis using Mobile Devices}.
\newblock In {\em Proceedings of the 29th Annual Computer Security Applications
  Conference (ACSAC)}, 2013.

\bibitem{SVOC13}
M.~Saini, S.~P. Venkatagiri, W.~T. Ooi, and M.~C. Chan.
\newblock The jiku mobile video dataset.
\newblock In {\em Proceedings of the 4th ACM Multimedia Systems Conference},
  MMSys '13, pages 108--113, New York, NY, USA, 2013. ACM.

\bibitem{DisturbingFake}
T.~Shelton.
\newblock {The most disturbing fake videos making the rounds in Syria}.
\newblock
  \url{http://www.globalpost.com/dispatch/news/regions/middle-east/syria/121109/fake-syria-videos-images},
  November 2012.

\bibitem{WBRN15}
H.~Wang, X.~Bao, R.~Roy~Choudhury, and S.~Nelakuditi.
\newblock Visually fingerprinting humans without face recognition.
\newblock In {\em Proceedings of the 13th Annual International Conference on
  Mobile Systems, Applications, and Services}, 2015.

\bibitem{zhang2012video}
W.~Zhang, R.~Zhang, X.~Liu, C.~Wu, and X.~Niu.
\newblock A video watermarking algorithm of h. 264/avc for content
  authentication.
\newblock {\em Journal of Networks}, 7(8):1150--1154, 2012.

\end{thebibliography}

\end{document}